\documentclass[11pt]{article}
\usepackage{amssymb}
\usepackage{amsmath}
\usepackage{graphics}
\usepackage{epsfig}
\usepackage{a4wide}
\usepackage{cite}
\usepackage{multirow}
\usepackage{color}
\usepackage{mathrsfs}
\usepackage{graphicx}

\DeclareGraphicsExtensions{.eps}
\begin{document}

\title{ P-wave excited $B_c^{**}$ meson photoproduction at the LHeC}
\author{ He Kai, Bi Huan-Yu\footnote{email:bihy@mail.ustc.edu.cn}, Zhang Ren-You, Li Xiao-Zhou and Ma Wen-Gan \\
\\
{\small State Key Laboratory of Particle Detection and Electronics, } \\
{\small University of Science and Technology of China, Hefei 230026, China} \\
{\small Department of Modern Physics, University of Science and Technology of China, Hefei 230026, China} }
\date{}
\maketitle \vskip 15mm
\maketitle
\vskip 15mm
\begin{abstract}
As an important sequential work of the S-wave $B_c^{(*)}$ ($^1S_0(^3S_1)$) meson production at the
large hadron electron collider (LHeC), we investigate the production of the P-wave excited
$B_c^{**}$ states ($^1P_1$ and $^3P_J$ with $J=0,1,2$) via photoproduction mechanism within
the framework of nonrelativistic QCD at the LHeC. Generally, the  $e^-+P \to \gamma+ g \rightarrow B_c^{**} +b+ \bar{c}$
process is considered as the main production mechanism at an electron-proton collider due to the
large luminosity of the gluon. However, according to our experience on the S-wave
$B_c^{(*)}$ meson production at the LHeC, the extrinsic production mechanism, i.e.,
$e^-+P \to\gamma+c \rightarrow B_c^{**} +b$ and $e^-+P \to\gamma+\bar{b} \rightarrow B_c^{**}+\bar{c}$, could also
provide dominating contributions at low $p_T$ region. A careful treatment between these channels
is performed and the results on total and differential cross sections, together with main uncertainties are discussed. Taking the quark masses $m_b=4.90\pm0.40$ GeV and $m_c=1.50\pm0.20$ GeV into account and summing up all the production channels, we expect to accumulate
$(2.48^{+3.55}_{-1.75}) \times 10^4$ $B_c^{**}({^{1}P_1})$,
$(1.14^{+1.49}_{-0.82}) \times 10^4$  $B_c^{**}({^{3}P_0})$,
$(2.38^{+3.39}_{-1.74}) \times 10^4$  $B_c^{**}({^{3}P_1})$ and
$(5.59^{+7.84}_{-3.93}) \times 10^4$  $B_c^{**}({^{3}P_2})$ events
at the $\sqrt{S}=1.30~\rm{TeV}$ LHeC in one operation year with luminosity ${\cal L}= 10^{33}$ cm$^{-2}$s$^{-1}$.
With such sizable events, it is worth studying the properties of excited
P-wave $B_c^{**}$ states at the LHeC.
\end{abstract}

\vskip 35mm

\vfill \eject
\baselineskip=0.32in
\makeatletter      
\@addtoreset{equation}{section}
\makeatother       
\vskip 5mm

\renewcommand{\theequation}{\arabic{section}.\arabic{equation}}
\renewcommand{\thesection}{\Roman{section}.}
\newcommand{\nb}{\nonumber}


\vskip 5mm
\section{INTRODUCTION}

\par
The doubly heavy meson physics has aroused great interest due to its nature, which can be studied in the framework of nonrelativistic QCD (NRQCD)\cite{NRQCD}. The production of the doubly heavy meson can be factorized into  a hard production of two heavy quark pairs which can be described by perturbation QCD (pQCD), and a soft term related to the nonpertubative binding of them. Thus, it is a good laboratory for testing NRQCD, pQCD and QCD potential models.

\par
Among the doubly heavy mesons, the $B_c$ meson\footnote{$^{1}S_{0}$ $B_c$ state is denoted as ground state $B_c$, $^{3}S_1$ $B_c$ state is denoted as $B_c^{*}$, and four P-wave ($^{1}P_1$, $^{3}P_0$, $^{3}P_1$ and $^{3}P_2$) $B_c$ states are denoted as $B_c^{**}$. } is especially interesting for its unique properties, which is the only  observed  meson  composed  of  a heavy quark and a heavy antiquark of different flavors. Unlike the charmonium and bottomonium states, which have `hidden flavor', the $B_c$ meson is made of a charm quark and a bottom antiquark, and the production of $B_c$ meson must be accompanied by additional heavy quarks especially in hadronic production \cite{heavy-quark, Brambilla:2010cs}. For example, production of a color-singlet $(c\bar{c})_1$ and a color-octet $(c\bar{c})_8$ quarkonium states are allowed for the channel $\gamma+g \rightarrow |(c\bar{c})_{1/8}\rangle +g/\gamma$. Therefore, compared with the production cross sections of hidden flavor quarkonia, the cross section of $B_c$ meson is  suppressed by not only  the phase space but also the higher order in the coupling constants of leading-order diagrams. Due to the small production rate and the low colliding luminosity, the $B_c$ meson was not found at LEP despite of careful searches \cite{Ackerstaff:1998zf, Abreu:1996nz, Barate:1997kk}. At hadron colliders, the background is extremely serious. With much time and effort, the ground state $B_c$ meson was finally observed by the CDF at Tevatron in 1998 \cite{Abe:1998wi, Abe:1998fb}. Now we also need more data to understand the properties of $B_c$ meson, such as mass spectrum, lifetime and decay. Thus, the researches on various production mechanisms at high energy colliders are required to study the properties of $B_c$ meson.

\par
The hadronic production of $B_c$ meson was amply studied directly via gluon-gluon fusion and etc, or indirectly via  top quark, $W$ or Higgs boson decays \cite{prod1, prod2, prod3, prod4, prod5, prod6, prod7,prod8,prod9,prod10,prod100,prod11,prod12,prod13,prod14,prod15,prod16,prod17,prod18,prod19,prod20,prod26,Jiang:2015pah}. These studies indicated that at a hadron collider, such as the Large Hadron Collider (LHC) or Tevatron, sizable $B_c$ events can be produced due to the powerful colliding energy and high luminosity. The $B_c$ meson production at electron-positron colliders, such as super $Z$-factory and international linear collider were discussed in \cite{prod21,prod22,prod23,prod24,prod25, prod27, Yang:2010yg,Berezhnoy:2016etd}. These lepton platforms have more clean background, hence are more suitable for precision measurement. For example, the authors in \cite{prod21, prod27} are interested in the forward-backward asymmetry in the production of doubly heavy-flavored hadron at the $Z$-factory. The electron-proton collider, which combines the advantages of hadron collider and lepton collider, may provide good opportunity to study doubly heavy-flavored hadrons. Thus, we have studied the $B_c^{(*)}$ meson and doubly heavy baryon production \cite{ep-S,Huan-Yu:2017emk} at the large hadron electron collider (LHeC) \cite{AbelleiraFernandez:2012cc} and future circular collider-based electron-proton collider (FCC-$ep$), and we found these colliders are very helpful for study doubly heavy-flavored hadrons.

\par
Recently, a new state has been observed by the ATLAS experiment \cite{bc-s-lhc}, whose mass and decay mode are consistent with the theoretical prediction of the second S-wave state $B_c^{\pm}(2S)$. The $B_c^{\pm}$ state is reconstructed through its decay to the ground state accompanied with two oppositely charged pions, and the $B_c$ ground state is detected through its decay $B_c^{\pm} \to J/\psi \pi^{\pm}$. Besides, the excited $B_c$ states can also decay (or in a cascade way) to the ground state through the electric or magnetic dipole transitions. In contrast with the hadronic decay, the feature of the electromagnetic decay of the excited $B_c$ states is that the characteristic product is an additional photon with energy about dozens or hundreds of MeV \cite{Monteiro:2016rzi,Monteiro:2016ijw,Wang:2015yea} rather than pion. The measurements to the characteristic products, i.e., the pion or photon, can be treated as the signals for the discovery of the excited $B_c$ states and the measurements of the ground or excited $B_c$ states can provide the opportunity for extracting information on the mass spectrum of the $(c\bar{b})$ bound states and QCD potential models.

\par
Generally speaking, the  excited $B_c$ states shall decay (or in a cascade way) to the ground state via  electromagnetic or hadronic interactions with almost $100\%$ probability, since $B_c$ carry both $b$ and $c$ flavour-quantum number. Besides, its excited states may not be discriminated easily from its ground state in experiments\cite{Brambilla:2010cs}. Thus, it's necessary to estimate the production rate of excited P-wave $B_c^{**}$ states, which will contribute to the production rate of S-wave $B_c^{(*)}$ states. On the other hand, it is helpful for the discovery of the excited $B_c$ states to give the dynamic distributions of the production. Studies on the production of excited $B_c$ states have been done in the literature\cite{prod10, prod12, prod21, prod22, Yang:2010yg,prod100}, and they found that the excited P-wave $B_c^{**}$ states can provide about $14\%$ $\sim$ $17\%$ contributions compared to the S-wave $B_c^{(*)}$ states.

\par
As indicated in \cite{ep-S}, large number of $B_c^{(*)}$ mesons (about $6\times10^5$ events per year) can be produced at LHeC. Motivated by the discovery of the second S-wave state $B_c^{\pm}$ and the sizable $B_c^{(*)}$ events at the LHeC, we are interested in whether enough P-wave $B_c^{**}$ events can be accumulated at the LHeC. In this paper, in addition to gluon-induced channel $\gamma + g \to B_c^{**}+b+\bar{c}$, two extrinsic heavy quark channels $\gamma + c\to B_c^{**}+b$ and $\gamma +\bar{b}\to B_c^{**}+\bar{c}$ are included. Although the density of $\bar{b}$ and $c$ quarks are small in proton, the contributions of $\gamma + c$ and $\gamma + \bar{b}$ channels cannot be neglected for the larger phase space and lower order in the coupling constants compared to the  $\gamma + g$ channel.

\par
The photoproduction of the  $B_c^{**}$ meson at the LHeC can be divided into three steps, which contain three subprocesses,
\begin{eqnarray}\label{asdf}
\label{eq:lo_processes}
&&{e^-}+{P} \to \gamma+g \to  (c\bar{b})[n]+b+\bar{c} \to B_c^{**} + b+ \bar{c} ,\nonumber \\
&&{e^-}+{P} \to\gamma+c \to (c\bar{b})[n]+b \to B_c^{**} + b ,\nonumber \\
&&{e^-}+{P} \to\gamma+\bar{b} \to (c\bar{b})[n]+\bar{c} \to B_c^{**} + \bar{c}.
\end{eqnarray}
First, the photon beams are produced from the electron bremsstrahlung and the partons are radiated from the protons. The density of  photon beams  can be described by the Weizs$\ddot{a}$cker-Williams approximation (WWA)\cite{WWA}, and the parton densities are described by the parton distribution functions (PDFs). Second, photon beams interact with the partons and a diquark state with certain quantum numbers $(c\bar{b})[n]$ is produced, and this step can be calculated by the pQCD. Finally, the $(c\bar{b})[n]$ are bounded together to form the $B_c^{**}$ meson through nonperturbative effect, which can be described by the nonperturbative matrix element, and the matrix element is proportional to the inclusive transition probability of the $(c\bar{b})[n]$ diquark to the bound state $B_c^{**}$. 
In this work, we only focus on four color-signet  diquark states of $(c\bar{b})[n]$, i.e., $(c\bar{b})_1[{^{1}P_1}]$, $(c\bar{b})_1 [{^{3}P_{0}]}$, $(c\bar{b})_1[{^{3}P_1]}$, and $(c\bar{b})_1 [{^{3}P_{2}]}$\footnote{We estimate the color-octet $|(c\bar{b})_8[^1S_0/^3S_1]g\rangle$ contribution to the P-wave $B_c^{**}$ meson production and find this contribution is small (about 4\% of the color-signet contribution), thus we neglect it in the following discussion.}.
\par
This paper is organized as follows: we present calculation details in Section II. The numerical results are given in Section III  and the summary is presented in Section IV.


\vskip 5mm
\section{OUTLINE OF THE CALCULATION}

\par
Based on pQCD, the total cross section of the $B_c^{**}$ meson production can be factorized into the convolution of the parton/photon density functions and the partonic cross section $d\hat{\sigma}_{\gamma i}(\mu,x_1,x_2)$ as follows:
\begin{eqnarray}
\label{eq:diff}
d \sigma(e^- + P \to B_c^{**} + X) = \int dx_1 dx_2 \sum_{i=c,\bar{b},g} f_{\gamma/e^-}(x_1)
f_{i/P}(\mu,x_2) d\hat{\sigma}_{\gamma i}(\mu,x_1,x_2),
\end{eqnarray}
here we have taken the renormalization scale $\mu_r$ and the factorization scale $\mu_f$ to be the same, i.e., $\mu_r=\mu_f=\mu$. $f_{i/P}$ are the PDFs and $f_{\gamma/e^-}$ is the photon density function which is described by the WWA as
\begin{eqnarray}
\label{eq:WWA}
f_{\gamma/e^-}(x) = \frac{\alpha}{2 \pi} \bigg [ \frac{1+(1-x)^2}{x} {\rm ln} \frac{Q^2_{\rm max}}{Q^2_{\rm min}} + 2 m_{e}^2 x \bigg (\frac{1}{Q^2_{\rm max}} -\frac{1}{Q^2_{\rm min}}\bigg) \bigg ],
\end{eqnarray}
where $x$ is the fraction of the longitudinal momentum of the photon to electron beams. $Q^2_{\rm {min}}$ and $Q^2_{\rm {max}}$ are the minimum and maximum photon virtuality which can be expressed as
\begin{eqnarray}
\label{eq:Q^2}
Q^2_{\rm min} &=& \frac{m_{e}^2x^2}{1-x}, \nonumber\\
Q^2_{\rm max} &=& (\theta_c E_{e})^2(1-x)+Q^2_{\rm min},
\end{eqnarray}
where $\theta_c$ is the electron-scattering angle and $E_e$ is the the electron beam energy, which are determined by the collider. In this work we set $\theta_c = 32$ mrad which is consistent with the choices in Refs.\cite{bbb,bbbb} and is satisfied with the requirement of $\theta_c\ll1$ rad \cite{Klasen:1997br,Klasen:2002xb}.

\par
To avoid the `double counting' between $\gamma + g$ and $\gamma + q$ channels, the general-mass variable-flavor-number  scheme (GM-VFNs) \cite{Olness:1997yc, Aivazis:1993kh, Aivazis:1993pi, Amundson:2000vg, Kniehl:2005mk} is adopted here. The cross section under the GM-VFNs is
\begin{eqnarray}
\label{eq:total}
d \sigma (e^- + P \to B_c^{**} + X )&=& f_{\gamma/e^-}(x_1)  f_{g/P}(\mu,x_2)
                  \otimes d\tilde{\sigma}_{\gamma g}(\mu,x_1,x_2) + \nonumber\\
       && \sum_{q=c,\bar{b}} f_{\gamma/e^-}(x_1) [ f_{q/P}(\mu,x_2) -
          f^{sub}_{q/P}(\mu,x_2) ] \otimes d\tilde{\sigma}_{\gamma q}(\mu,x_1,x_2). \nonumber\\
\end{eqnarray}
$d\tilde{\sigma}_{\gamma g}$ contains mass logarithmic terms $\ln(Q^2/m^2_q)$, and these logarithmic terms originate in the  Feynman diagrams which contains initial gluon splitting to a heavy quark pair $g \to q\bar{q}$. $d{\sigma}$ is the infrared-safe partonic cross section which avoid the logarithmic terms through the subtraction of the term $f^{sub}_{q/P}(\mu,x_2)$:
\begin{eqnarray}
\label{eq:sub-1}
f^{sub}_{q/P}(\mu,x_2) = \int_{x_2}^{1} f_{g/P}(\frac{x_2}{y}) \frac{\alpha_s(\mu)}{2\pi} {\rm{ln}} \frac{\mu^2}{m_q^2}P_{g\to q}(y) \frac{dy}{y},
\end{eqnarray}
where $P_{g\to q}(y) = \frac{1}{2} (1 - 2y +2y^2)$ is the $g \to q \bar{q}$ splitting function.


\par
The partonic hard cross section $d\tilde{\sigma}_{\gamma i}$ can also be factorized into a diquark production $(c\bar{b})[n]$ multiply by the nonperturbative matrix element $\langle \mathcal{O}^{B_c^{**}}  \rangle$,
\begin{eqnarray}
\label{eq:partonic}
d\tilde{\sigma}_{\gamma i}= \frac{\langle \mathcal{O}^{B_c^{**}}  \rangle}{4 E_{\gamma}E_{i}|\vec{v}_{\gamma}-\vec{v}_{i}|}{\bf \overline{\sum} }|\mathcal{M}|^2 d\Phi.
\end{eqnarray}
For the color-singlet $B_c^{**}$ meson production, $\langle \mathcal{O}^{B_c^{**}}  \rangle$ is related to the  first derivative of the  wave function  at the origin of the  $(c\bar{b})[n]$ bound state \cite{NRQCD}, i.e., $\langle \mathcal{O}^{B_c^{**}}  \rangle \simeq |R'_{P}(0)|^2/(4\pi)$, and  $R'_{P}(0)$ can be calculated from the potential model \cite{Buchmuller:1980su, Eichten:1995ch, Eichten:1994gt}. $d\Phi$ stands for the phase-space element and $\mathcal{M}$ for the hard-scattering amplitudes of $(c\bar{b})[n]$ production which contains the precise spin and orbit information,
\begin{eqnarray}
\label{partoniceq1}
{\mathcal{M}^{S=0,L=1}} &=& \varepsilon_{\alpha} (p_3) \frac{d}{dq_{\alpha}}  T \vert_{q=0}, \\
\label{partoniceq2}
{\mathcal{M}^{S=1,L=1}} &=& \varepsilon_{\alpha \beta}^J (p_3) \frac{d}{dq_{\alpha}}  T \vert_{q=0}.
\end{eqnarray}
where $T$ is related to the Feynman diagrams and  $q$ is the relative momentum among the quarks inside $(c\bar{b})[n]$, which can be set as zero after the derivation of the amplitudes. $\varepsilon_{\alpha} (p_3)$ is the polarization vector of the angular momentum triplet $^1P_1$ diquark and $\varepsilon_{\alpha \beta}^J (p_3)$ stands for the polarization tensor of the spin-triplet P-wave $^3P_J$ diquark. The summation over the polarizations obey the following relations:
\begin{eqnarray}
\sum_{\rm{polarizations}} \varepsilon_{\alpha} \varepsilon^{*}_{\alpha'} &=& \Pi_{\alpha \alpha'}, \nonumber \\
\varepsilon_{\alpha \beta}^{0} \varepsilon_{\alpha' \beta'}^{0*} &=& \frac{1}{3} \Pi_{\alpha \beta} \Pi_{\alpha' \beta'}, \nonumber \\
\sum_{\rm{polarizations}} \varepsilon_{\alpha \beta}^{1} \varepsilon_{\alpha' \beta'}^{1*} &=& \frac{1}{2} (\Pi_{\alpha \alpha'} \Pi_{\beta \beta'}- \Pi_{\alpha \beta'}\Pi_{\alpha' \beta}), \nonumber \\
\sum_{\rm{polarizations}} \varepsilon_{\alpha \beta}^{2} \varepsilon_{\alpha' \beta'}^{2*} &=& \frac{1}{2} (\Pi_{\alpha \alpha'} \Pi_{\beta \beta'}+ \Pi_{\alpha \beta'}\Pi_{\alpha' \beta}) -\frac{1}{3} \Pi_{\alpha \beta} \Pi_{\alpha' \beta'},
\end{eqnarray}
where
\begin{eqnarray}
\Pi_{\alpha \beta} = -g_{\alpha \beta} + \frac{p_{3 \alpha} p_{3 \beta}}{M^2},
\end{eqnarray}
and $M$ is the mass of $B^{**}_c$ meson.
\par
The Feynman diagrams and amplitudes are generated by FeynArts\cite{FA} and are provided in Appendix. Further simplification on the amplitudes are handled by FeynCalc\cite{FeynC} and FeynCalcFormLink\cite{FCF}. Numerical calculations are  performed by FormCalc\cite{FC}.

\vskip 5mm
\section{NUMERICAL RESULTS}

\par
The derivative of the wave function at the origin $\vert R'_P(0) ~\vert^2=0.201~ {\rm GeV^5}$ is
taken from Ref.\cite{Eichten:1994gt}. The P-wave $B_c^{**}$ mesons mass $M$ is taken as same
with the S-wave $B_c^{(*)}$ mesons, which is the requirement for the NRQCD formalism and are
explained in Ref.\cite{prod10}, i.e., $M=m_b+m_c$ with $b$-quark mass $m_b=4.90 ~{\rm GeV}$ and
$c$-quark mass $m_c=1.50 ~{\rm GeV}$\cite{Eichten:1994gt}. The electron mass $m_e$ is taken as $0.51\times10^{-3}$ GeV
and the fine-structure constant is chosen as $\alpha=1/137$. The renormalization and factorization
scale are set to be the transverse mass of the $B_c^{**}$ meson
$\mu=\mu_r=\mu_f=M_T = \sqrt{p_T^2+M^2}$. We use CT10nlo\cite{ct10nlo} as default and the $\alpha_s$ is extracted from the PDFs.

\par
The cross sections for all the production channels with four collision energies at two
electron-proton colliders are presented in Table \ref{total-cs}, i.e., for the LHeC
$\sqrt{S} = 1.30, 1.98~\rm{TeV}$ \cite{AbelleiraFernandez:2012cc} which corresponds to two
beam energy sets designs as $E_{e} = 60,~140 ~\rm{GeV}$ and $E_{P} = 7~\rm{TeV}$. For the FCC-$ep$
we take $\sqrt{S} = 7.07, 10.00~\rm{TeV}$ \cite{Acar:2016rde} which correspond to two beam energy
sets as $E_{e} = 250,~500 ~\rm{GeV}$ and $E_{P} = 50~\rm{TeV}$ separately. Here, we use
$\sigma_{\gamma g}$, $\sigma_{\gamma c}$ and $\sigma_{\gamma \bar{b}}$ to denote  the cross
sections of $\gamma + g$, $\gamma + c$ and $\gamma + \bar{b}$ channels, respectively. Summing
up all the contributions of three channels and four P-wave $B_c^{**}$ states, we find that the
contribution from P-wave $B_c^{**}$ states can be $19.7\%$\footnote{Considering the main uncertainties from quark masses, 
this ratio should be $19.7^{+6.82}_{-8.76}\%$ at the LHeC with $\sqrt{S} = 1.30$ TeV.}, $19.9\%$, $19.8\%$ and $20.2\%$
of the S-wave $B_c^{(*)}$ state production (59.01, 95.91, 296.75 and 399.05 pb) \cite{ep-S} for above four colliding energies cases.
That shows these ratios are larger than the corresponding ones at the $Z$-factory, LHC
($\sqrt{S}=14$ TeV) and Tevatron ($\sqrt{S}=1.96$ TeV), where the ratios are about $17.40\%$,
$16.20\%$ and $14.93\%$, respectively \cite{prod10, prod21}. We can conclude that the LHeC and
FCC-$ep$ colliding experiments might have advantages on the study of the P-wave $B_c^{**}$ states,
thus it is worth to study the excited states at these two colliders. By summing up the four P-wave
$B_c^{**}$ states, We see also that the $\gamma + \bar{b}$ channel makes the largest contributions
to the P-wave $B_c^{**}$ state production, while the contributions from the $\gamma + g$
and $\gamma + c$ channels are at the same order which only provide about $5\%$ and $1\%$ to the
total cross section for various electron-proton colliding energies. Besides, by summing up all the
production channels, the four $B_c^{**}$ states, i.e., $^1P_1$, $^3P_0$, $^3P_1$ and $^3P_2$,
provide individually about $21\%$, $10\%$, $20\%$ and $48\%$ contributions to the total cross
section, respectively. The total cross section are similar at different colliding energies, and
in the following, we focus on the  $B_c^{**}$ meson production at the $\sqrt{S} = 1.30 ~\rm{TeV}$
LHeC.
\begin{table}[htb]
\caption{The cross sections (pb) for $B_c^{**}$ production at two electron-proton colliders. Four electron-proton colliding energies are adopted, i.e., $\sqrt{S}$=1.30 and 1.98 TeV for LHeC, and $\sqrt{S}$=7.07 and 10.0 TeV for FCC-$ep$}
\begin{center}
\resizebox{\textwidth}{!}{
\begin{tabular}{c|c|cccc|cccc|cccc|c}
\hline
  &\multirow{2}{*}{$\sqrt{S}$} & \multicolumn{4}{|c|}{$\sigma_{\gamma c}$}  & \multicolumn{4}{|c|}{$\sigma_{\gamma \bar{b}}$}  & \multicolumn{4}{|c|}{$\sigma_{\gamma g}$} & \multirow{2}{*}{Total} \\ \cline{3-14}
        &       & $^1P_1$ & $^3P_0$ & $^3P_1$ & $^3P_2$ & $^1P_1$ & $^3P_0$ & $^3P_1$ & $^3P_2$ & $^1P_1$ & $^3P_0$ & $^3P_1$ & $^3P_2$ & \\ \hline
\multirow{2}{*}{LHeC} &  1.30  & $4.87\times10^{-2}$ & $1.28\times10^{-2}$ & $2.68\times10^{-2}$ & $7.13\times10^{-2}$ & 2.30 & 1.10 & 2.27 & 5.22 & $1.27\times10^{-1}$ & $3.08\times10^{-2}$ & $7.72\times10^{-2}$ & $2.99\times10^{-1}$ & 11.6 \\
 &  1.98  & $7.26\times10^{-2}$  & $1.94\times10^{-2}$ & $4.05\times10^{-2}$ & $1.07\times10^{-1}$ & 3.80 & 1.81 & 3.75 & 8.61 & $2.10\times10^{-1}$ & $5.06\times10^{-2}$ & $1.28\times10^{-1}$ & $4.90\times10^{-1}$ & 19.1 \\
\multirow{2}{*}{FCC-$ep$} & 7.07 & $1.68\times10^{-1}$ & $4.66\times10^{-2}$ & $9.64\times10^{-2}$ & $2.53\times10^{-1}$ & 11.8 & 5.55 & 11.6 & 26.7 & $6.34\times10^{-1}$ & $1.52\times10^{-1}$ & $3.93\times10^{-1}$ & 1.46 & 58.8 \\
 & 10.0  & $2.14\times10^{-1}$ & $5.97\times10^{-2}$ & $1.23\times10^{-1}$ & $3.22\times10^{-1}$ & 16.3 & 7.63 & 15.9 & 36.7 & $8.65\times10^{-1}$ & $2.06\times10^{-1}$ & $5.35\times10^{-1}$ & 1.98 & 80.8 \\
\hline
\end{tabular} }
\label{total-cs}
\end{center}
\end{table}

\par
\begin{figure}[htbp]
\includegraphics[scale=0.55]{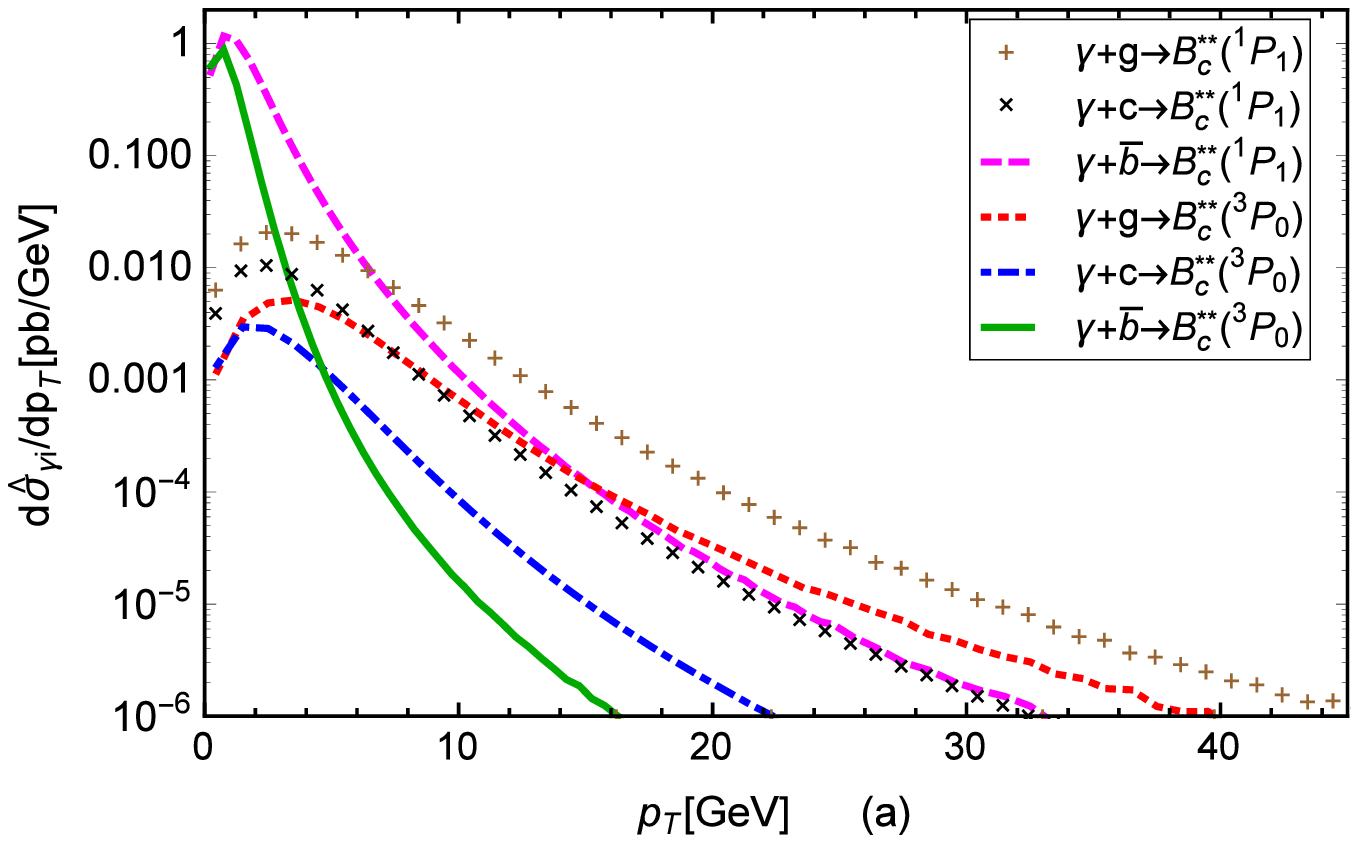}
\hspace{0in}%
\includegraphics[scale=0.55]{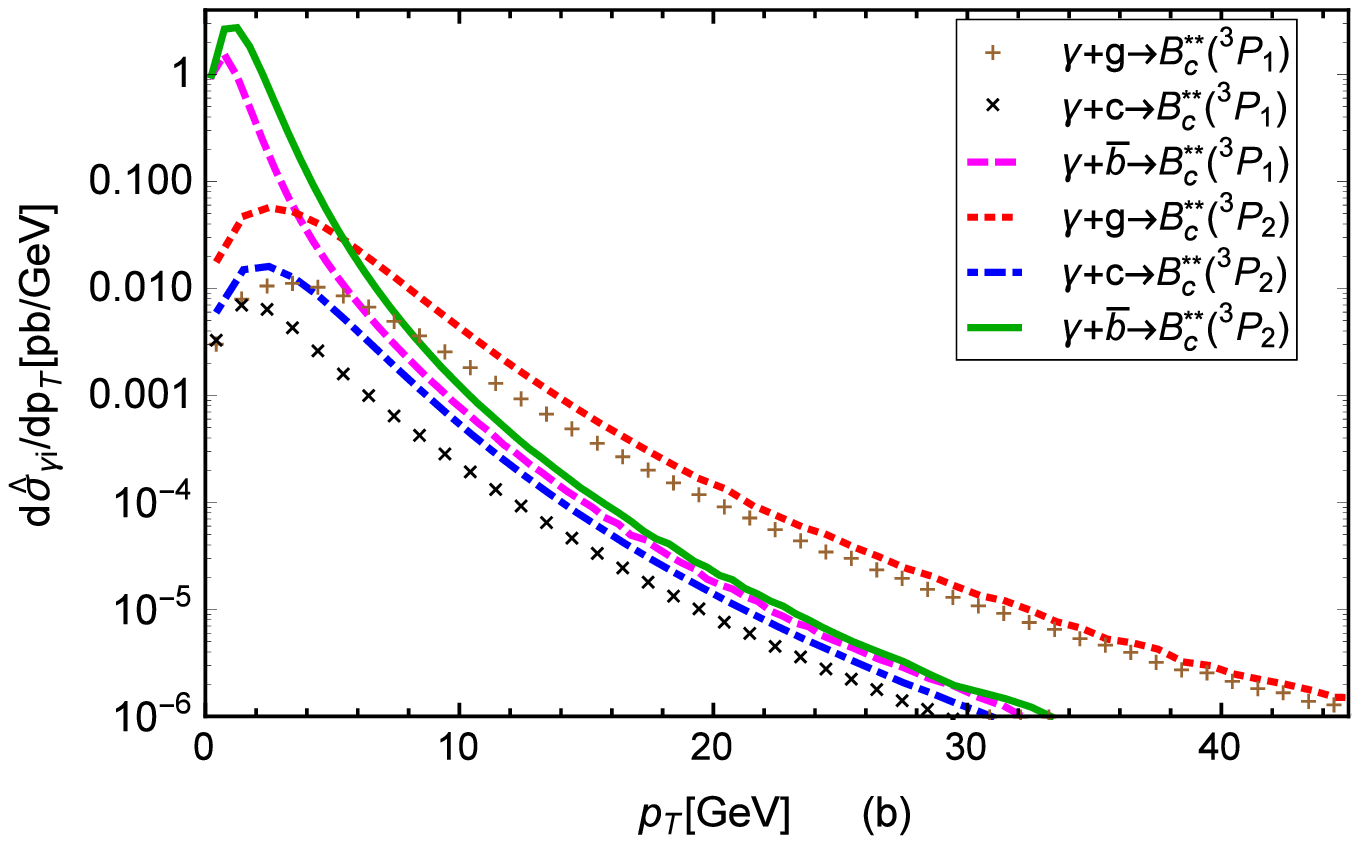}
\hspace{0in}%
\caption{The transverse momentum distributions for the $B_c^{**}$ meson at the
$\sqrt{S} = 1.30 ~\rm{TeV}$ LHeC. (a) For the processes of $e^-+P \to \gamma +i \to B_c^{**}(^1P_1/^3P_0)$. (b) For the processes of $e^-+P \to \gamma +i \to B_c^{**}(^3P_1/^3P_2)$.}
\label{pt1}
\end{figure}

The $B_c^{**}$ transverse momentum ($p_T$) distributions of all the production channels at the LHeC are
shown in Figs.\ref{pt1}(a,b). From the figures we can see that the contribution from
$\gamma + \bar{b}$ channel ominates in the low $p_T$ regions, but drops down more drastically than
that from the $\gamma + g$ channel. Although the contribution of the $\gamma + g$ channel is suppressed
in the low $p_T$ region, it becomes the main contribution when the $p_T$ goes up to large value range.
We can see if one wants to select the P-wave $B_c^{**}$ production signal from the $\gamma + g$
production channel at the LHeC, one can simply accept the events by imposing proper high lower
$p_T^{\rm lower}$ cut ($p_T > p_T^{\rm lower}$), and then the heavy quark initiated P-wave $B_c^{**}$
production events can be suppressed. We find also from both Figs.\ref{pt1}(a) and (b) that each
differential cross section curve for the P-wave $B_c^{**}$ meson from the $\gamma + \bar{b}$
production channel, has a pick around 1 GeV, which is quantitatively exceeded one or two
order of the those for the $\gamma + g$ and $\gamma + c$ production channels. Thus,
if the $p_T$ acceptance range of the $B_c^{**}$ mesons is limited all around 1 GeV, it is
advantageous for the investigation of the P-wave $B_c^{**}$ mesons from the $\gamma + \bar{b}$
production channel.

\par
\begin{figure}[htbp]
\includegraphics[scale=0.55]{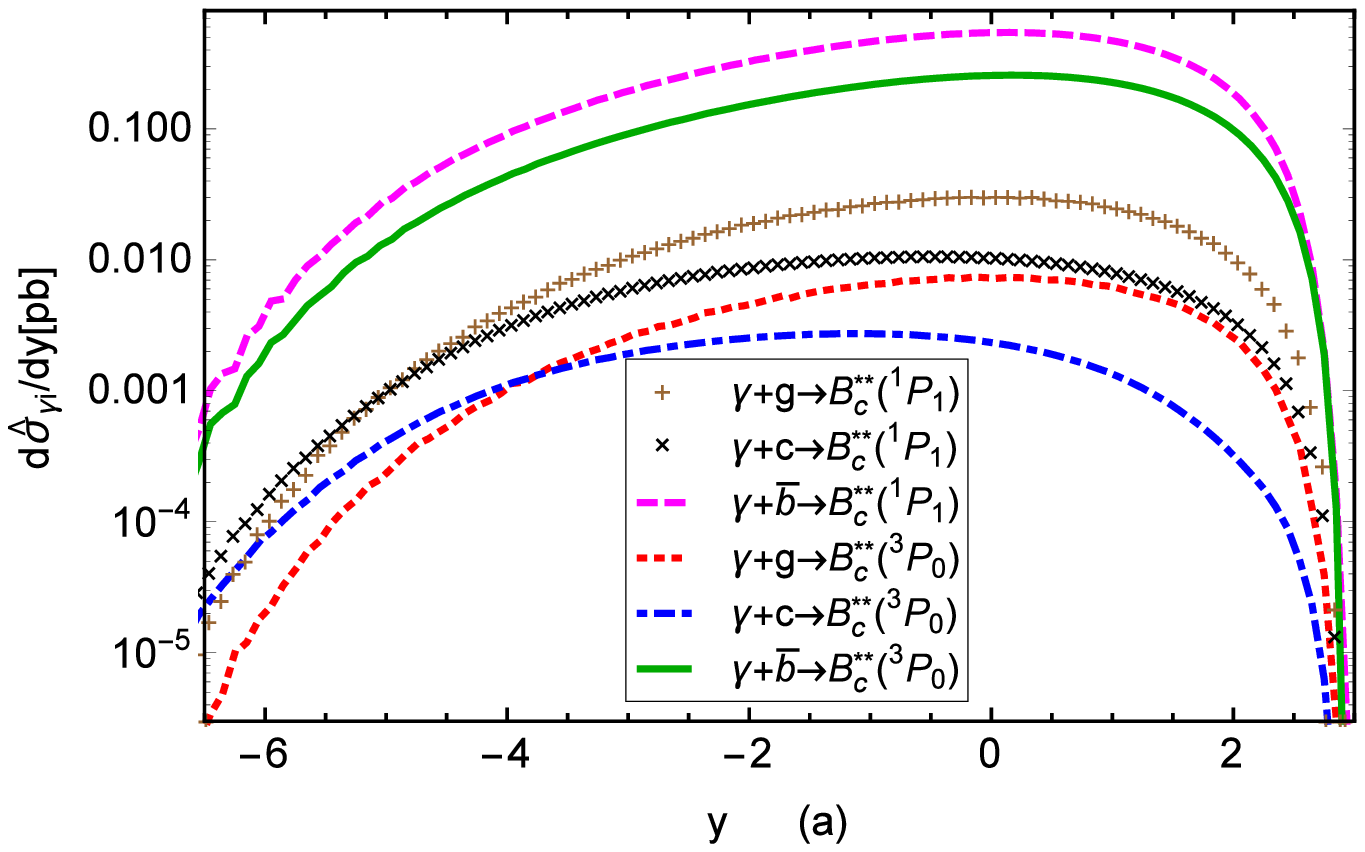}
\hspace{0in}%
\includegraphics[scale=0.55]{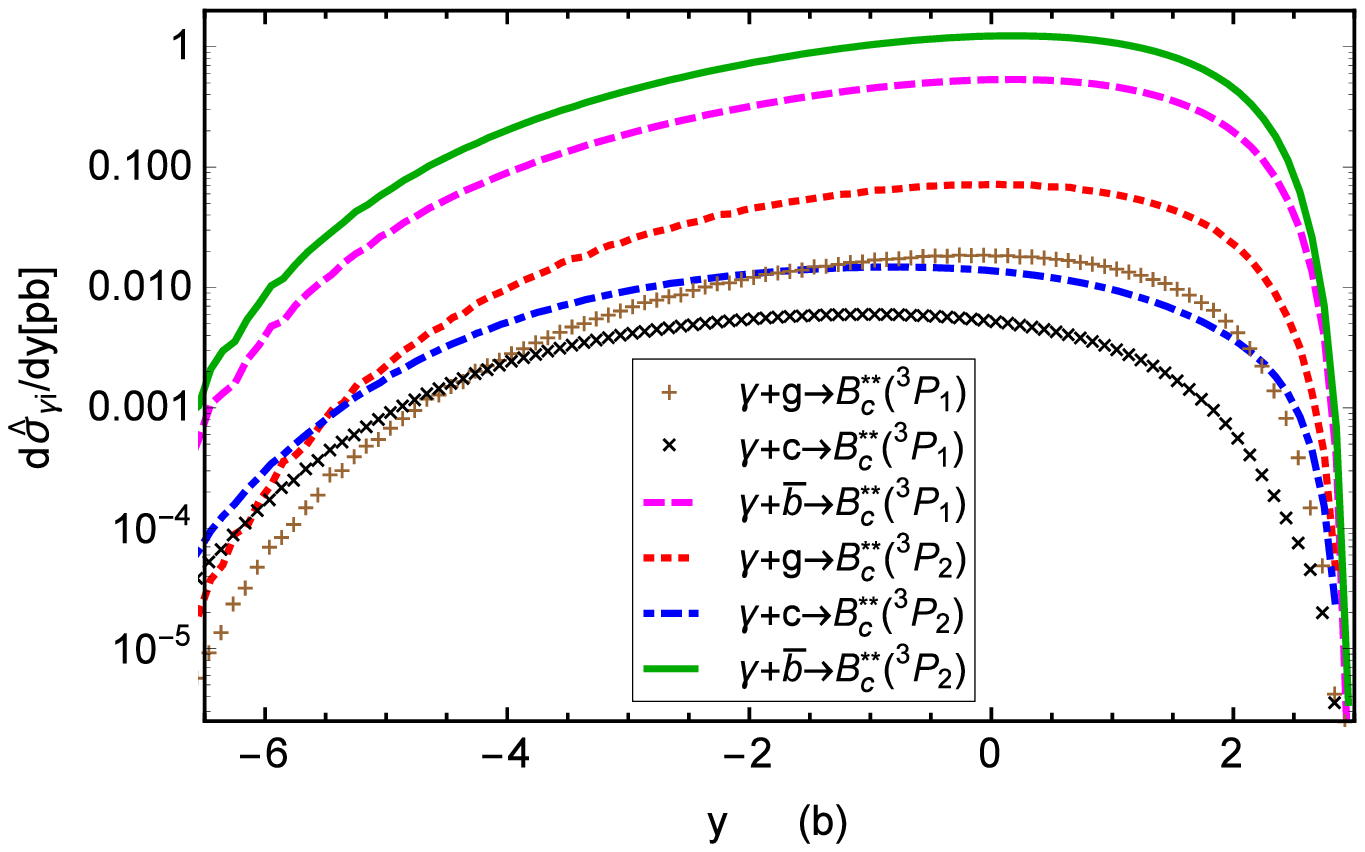}
\hspace{0in}%
\caption{The rapidity distributions for the $B_c^{**}$ meson at the
$\sqrt{S} = 1.30 ~\rm{TeV}$ LHeC. (a) For the processes of $e^-+P \to \gamma +i \to B_c^{**}(^1P_1/^3P_0)$. (b) For the processes of $e^-+P \to \gamma +i \to B_c^{**}(^3P_1/^3P_2)$.}
\label{y-1}
\end{figure}

We present the rapidity ($y$) distributions of $B_c^{**}$ mesons in Figs.\ref{y-1}(a,b).
There the asymmetry rapidity distributions indicate that the dominate contributions are located in
the region around $y=1$, due to the colliding energies of the incoming particles being
not equal and the majority of the photons radiated from electron beams carrying less energies than
the partons in the protons. Figs.\ref{y-1}(a,b) show that the $B_c^{**}$ meson rapidity distributions
of all the three channels drop sharply when $y$ increases from 2 to 3, while go down gently
if $y$ decreases from -4.

\par
We know that the total cross section for $B_c^{**}$ production should be sensitive to various
experimential cuts, such as the $p_T$ and $y$ cuts on the final mesons. The cross sections by applying
different $p_T$ and $y$ cuts on $B_c^{**}$ mesons are presented in Table \ref{ptcuts} and
Table \ref{ycuts} separately. The data show that if the events are accepted with the condition of
$p_T>1.0 ~\rm{GeV}$, the cross section via $\gamma+\bar{b}$ channel is about one order larger than
that from the $\gamma + g$ channel, which can be demonstrate also from the $p_T$ distributions
shown in Figs.\ref{pt1}(a,b). By summing up the three channels and the four P-wave $B_c^{**}$
states, we can get the ratio ($\frac{\sigma_{\rm{cuts}}}{\sigma_{\rm{No~cuts}}}$) as
$59.1\%$, $9.31\%$, $2.81\%$ by applying $p_T$ cuts as $1.0 ~\rm{GeV}$, $3.0 ~\rm{GeV}$ and
$5.0 ~\rm{GeV}$, respectively.
\begin{center}
\begin{table}[htb]
\caption{The cross sections (pb) for the $B_c^{**}$ production at the $\sqrt{S} = 1.30 ~\rm{TeV}$ LHeC
under various $p_T$ cuts.}
\resizebox{\textwidth}{!}{
\begin{tabular}{c|cccc|cccc|cccc}
\hline
$ $ & \multicolumn{4}{|c|}{$p_T \ge 1.0 ~\rm{GeV}$} & \multicolumn{4}{|c|}{$p_T \ge 3.0 ~\rm{GeV}$} & \multicolumn{4}{|c}{$p_T \ge 5.0 ~\rm{GeV}$} \\ \cline{2-13}
$ $ & $^1P_1$ & $^3P_0$ & $^3P_1$ & $^3P_2$ & $^1P_1$ & $^3P_0$ & $^3P_1$ & $^3P_2$ & $^1P_1$ & $^3P_0$ & $^3P_1$ & $^3P_2$  \\ \hline
$\sigma_{\gamma c}$ & $4.49\times10^{-2}$ & $1.14\times10^{-2}$ & $2.37\times10^{-2}$ & $6.50\times10^{-2}$ & $2.60\times10^{-2}$ & $5.57\times10^{-3}$ & $1.10\times10^{-2}$ & $3.40\times10^{-2}$ & $1.16\times10^{-2}$ & $2.12\times10^{-3}$ & $4.38\times10^{-3}$ & $1.31\times10^{-2}$  \\
$\sigma_{\gamma \bar{b}} $ & 1.44 & $3.43\times10^{-1}$ & 1.01 & 3.40 & $2.06\times10^{-1}$ & $9.10\times10^{-3}$ & $1.04\times10^{-1}$ & $3.51\times10^{-1}$ & $4.15\times10^{-2}$ & $8.52\times10^{-4}$ & $2.34\times10^{-2}$ & $5.34\times10^{-2}$ \\
$\sigma_{\gamma g}$ & $1.21\times10^{-1}$ & $2.96\times10^{-2}$ & $7.41\times10^{-2}$ & $2.80\times10^{-1}$ & $8.31\times10^{-2}$ & $2.14\times10^{-2}$ & $5.54\times10^{-2}$ & $1.76\times10^{-1}$ & $4.56\times10^{-2}$ & $1.17\times10^{-2}$ & $3.37\times10^{-2}$ & $8.44\times10^{-2}$  \\ \hline
Total & \multicolumn{4}{|c|}{6.85} & \multicolumn{4}{|c|}{1.08} & \multicolumn{4}{|c}{$3.26\times10^{-1}$} \\
\hline
\end{tabular}
}
\label{ptcuts}
\end{table}
\end{center}
\begin{center}
\begin{table}[htb]
\caption{The cross sections (pb) for the $B_c^{**}$ production at the $\sqrt{S} = 1.30 ~\rm{TeV}$ LHeC
under various $y$ cuts.}
\resizebox{\textwidth}{!}{
\begin{tabular}{c|cccc|cccc|cccc}
\hline
$ $ & \multicolumn{4}{|c|}{$|y| \le 1.0$} & \multicolumn{4}{|c|}{$|y| \le 2.0$} & \multicolumn{4}{|c}{$|y| \le 3.0$} \\ \cline{2-13}
$ $ & $^1P_1$ & $^3P_0$ & $^3P_1$ & $^3P_2$ & $^1P_1$ & $^3P_0$ & $^3P_1$ & $^3P_2$ & $^1P_1$ & $^3P_0$ & $^3P_1$ & $^3P_2$  \\ \hline
$\sigma_{\gamma c}$ & $1.91\times10^{-2}$ & $4.47\times10^{-3}$ & $9.66\times10^{-3}$ & $2.65\times10^{-2}$ & $3.39\times10^{-2}$ & $7.93\times10^{-3}$ & $1.70\times10^{-2}$ & $4.72\times10^{-2}$ & $4.21\times10^{-2}$ & $1.03\times10^{-2}$ & $2.18\times10^{-2}$ & $6.00\times10^{-2}$  \\
$\sigma_{\gamma \bar{b}} $ & 1.04 & $4.89\times10^{-1}$ & 1.02 & 2.34 & 1.78 & $8.44\times10^{-1}$ & 1.75 & 4.04 & 2.09 & $9.97\times10^{-1}$ & 2.07 & 4.75 \\
$\sigma_{\gamma g}$ & $5.79\times10^{-2}$ & $1.40\times10^{-2}$ & $3.55\times10^{-2}$ & $1.37\times10^{-1}$ & $1.00\times10^{-1}$ & $2.41\times10^{-2}$ & $6.03\times10^{-2}$ & $2.34\times10^{-1}$ & $1.17\times10^{-1}$ & $2.84\times10^{-2}$ & $7.10\times10^{-2}$ & $2.76\times10^{-1}$  \\ \hline
Total & \multicolumn{4}{|c|}{5.19} & \multicolumn{4}{|c|}{8.94} & \multicolumn{4}{|c}{10.5} \\
\hline
\end{tabular}
}
\label{ycuts}
\end{table}
\end{center}

\par
\begin{figure}[htbp]
\includegraphics[scale=0.55]{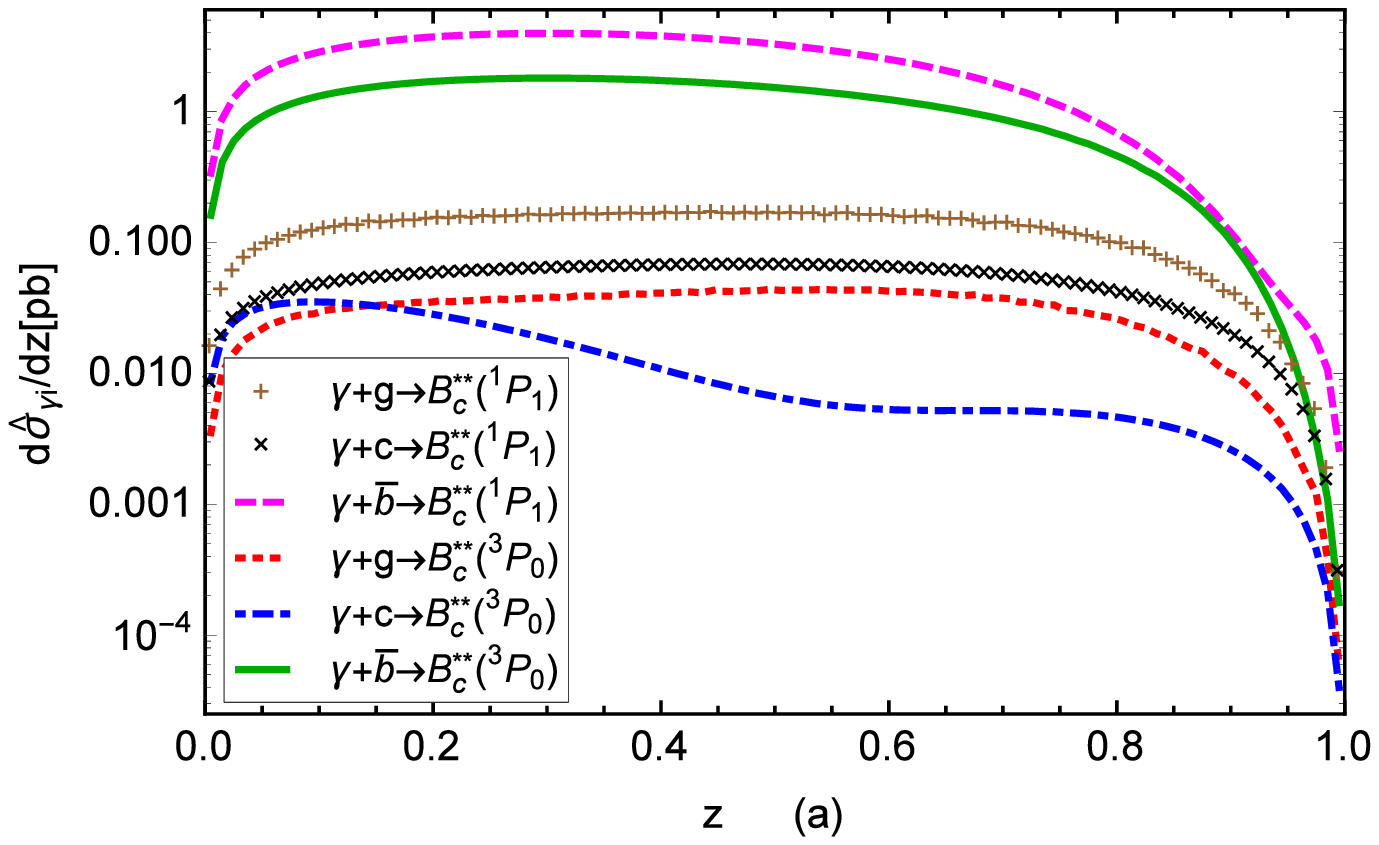}
\hspace{0in}
\includegraphics[scale=0.55]{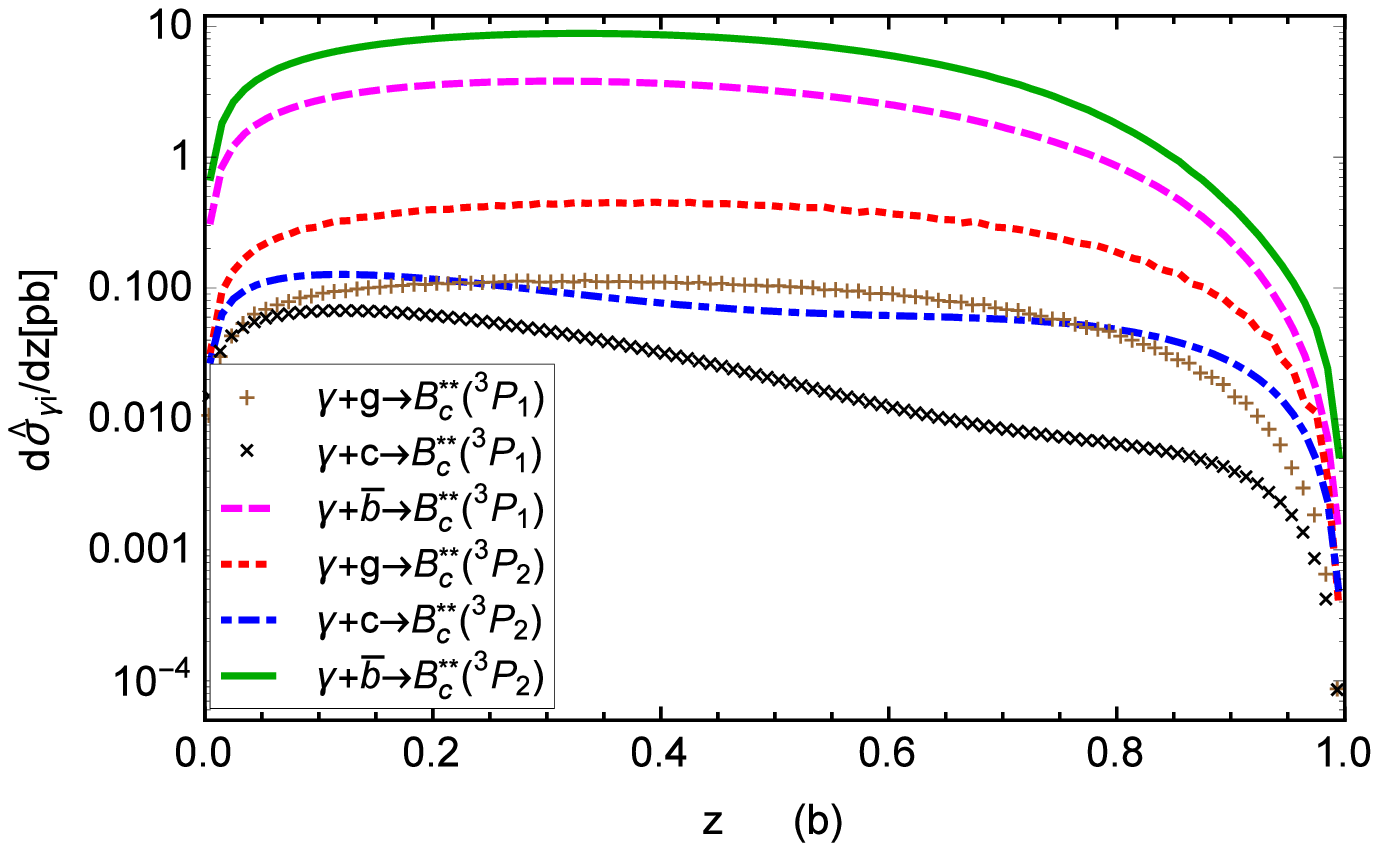}
\hspace{0in}
\caption{Differential cross sections $d\sigma/dz$ versus $z$ for the $B_c^{**}$ production at the
$\sqrt{S} = 1.30 ~\rm{TeV}$ LHeC. (a) For the processes of $e^-+P \to \gamma +i \to B_c^{**}(^1P_1/^3P_0)$. (b) For the processes of $e^-+P \to \gamma +i \to B_c^{**}(^3P_1/^3P_2)$.}
\label{z-1}
\end{figure}

In photoproduction experiments, the study on inelastic $B_c$ events can give information on the gluon
distribution in the nucleon \cite{cc}, and the inelasticity of the photoproduction can be denoted by the  variable $z = \frac{p_{B_c}\cdot p_P}{p_{\gamma} \cdot p_P}$, where $p_{B_c}$, $p_{\gamma}$ and $p_P$ denote the four-momenta of the $B_c$, $\gamma$ and proton, respectively. In the elastic domain, $z \approx 1$, and at low $z$ region, the resolved effect (the hadronic components could be radiated from the photon) should also make some contributions \cite{z-cuts-3}. Normally one can obtain clean samples of inelastic events in the range of $0.3\lesssim z \lesssim 0.9$ \cite{z-cuts-1,z-cuts-2,z-cuts-22, z-cuts-3}. Figs.\ref{z-1}(a,b) show the distributions of variable $z$, and in Table \ref{zcuts} we list the total cross sections at the $\sqrt{S} = 1.30 ~\rm{TeV}$ LHeC for different $B_c^{**}$ production processes by accepting the events in the $z$ range of $0.3 \le z \le 0.9$.
\begin{table}[htb]
\caption{The cross section (pb) in the range of $0.3 \le z \le 0.9$ for different $B_c^{**}$ production processes at the $\sqrt{S} = 1.30 ~\rm{TeV}$ LHeC.}
\begin{center}
\begin{tabular}{c|ccccc }
\hline
$ $ & $^1P_1$ & $^3P_0$ & $^3P_1$ & $^3P_2$ & Total  \\ \hline
$\sigma_{\gamma g}$        & $8.62\times10^{-2}$ & $2.15\times10^{-2}$ & $4.92\times10^{-2}$ & $1.99\times10^{-1}$ & $3.56\times10^{-1}$ \\
$\sigma_{\gamma \bar{b}} $ & 1.39 & $6.80\times10^{-1}$ & 1.40 & 3.26 & 6.73 \\
$\sigma_{\gamma c}$        & $3.34\times10^{-2}$ & $4.25\times10^{-3}$ & $1.01\times10^{-2}$ & $3.73\times10^{-2}$ & $8.51\times10^{-2}$ \\
\hline
\end{tabular}
\label{zcuts}
\end{center}
\end{table}

\par
We present the results for $\mu = 0.75{M_T}$, ${M_T}$ and
$1.25{M_T}$ in Table \ref{scale} separately. For the $\gamma + c$ and $\gamma + g$ production
channels, the cross sections decrease slightly when the scale becomes larger, while for the
$\gamma +\bar{b}$ channels, the situation is opposite. At fixed order, the scale dependence of the prediction would
lead to the uncertainties of final results. The scale uncertainties for the cross section
(defined as $\frac{\sigma(\mu=\mu')-\sigma(\mu=M_T)} {\sigma(\mu=M_T)}$ with $\mu'=0.75,1.25M_T$) via the
$\gamma + c$, $\gamma + \bar{b}$ and $\gamma + g$ channels
are $-6\%\sim 8\%$, $-79\%\sim 31\%$ and $-11\%\sim 15\%$, respectively. By summing up all the three channels and
the four P-wave $B_c^{**}$ states, the total cross section still increases with the increasement of
the scale owing to the large contributions from the $\gamma+ \bar{b}$ channels. Such a large scale
dependence could be reduced by involving higher-order QCD corrections. Furthermore, the renormalization scale dependence can 
be reduced by the QCD scale setting method
\cite{Wu:2013ei,Wu:2014iba}.
\begin{table}[htb]
\caption{The cross sections (pb) for the $B_c^{**}$ meson production at the $\sqrt{S} = 1.30 ~\rm{TeV}$
LHeC for $\mu = 0.75 M_T$, $M_T$ and $1.25 M_T$ separately.}
\begin{center}
\resizebox{\textwidth}{!}{
\begin{tabular}{c|cccc|cccc|cccc|c}
\hline
\multirow{2}{*}{$\mu$} & \multicolumn{4}{|c|}{$\sigma_{\gamma c}$}  & \multicolumn{4}{|c|}{$\sigma_{\gamma \bar{b}}$}  & \multicolumn{4}{|c|}{$\sigma_{\gamma g}$} & \multirow{2}{*}{Total} \\ \cline{2-13}
       & $^1P_1$ & $^3P_0$ & $^3P_1$ & $^3P_2$ & $^1P_1$ & $^3P_0$ & $^3P_1$ & $^3P_2$ & $^1P_1$ & $^3P_0$ & $^3P_1$ & $^3P_2$ & \\ \hline
$ 0.75M_T $ & $5.27\times10^{-2}$  & $1.38\times10^{-2}$ & $2.89\times10^{-2}$ & $7.70\times10^{-2}$ & $5.15\times10^{-1}$ & $2.08\times10^{-1}$ & $4.67\times10^{-1}$ & 1.12 & $1.46\times10^{-1}$ & $3.54\times10^{-2}$ & $8.90\times10^{-2}$ & $3.44\times10^{-1}$ & 3.10 \\
 $    M_T $ & $4.87\times10^{-2}$  & $1.28\times10^{-2}$ & $2.68\times10^{-2}$ & $7.13\times10^{-2}$ & 2.30 & 1.10 & 2.27 & 5.22 & $1.27\times10^{-1}$ & $3.08\times10^{-2}$ & $7.72\times10^{-2}$ & $2.99\times10^{-1}$ & 11.6 \\
$ 1.25M_T $ & $4.56\times10^{-2}$  & $1.20\times10^{-2}$ & $2.52\times10^{-2}$ & $6.69\times10^{-2}$ & 2.99 & 1.45 & 2.99 & 6.80 & $1.14\times10^{-1}$ & $2.78\times10^{-2}$ & $6.93\times10^{-2}$ & $2.68\times10^{-1}$ & 14.9 \\
\hline
\end{tabular} }
\label{scale}
\end{center}
\end{table}

\par
Finally, we discuss the quark mass dependence of the cross sections. We take
$m_c = 1.50\pm 0.20 ~\rm{GeV}$ and $ m_b = 4.90 \pm 0.40 ~\rm{GeV}$ into account. The $m_c$ is fixed
as its center values when discussing the uncertainty from $ m_b = 4.90 \pm 0.40 ~\rm{GeV}$ and vice
versa. The cross sections under different value of $m_c$ and $m_b$ are presented in
Table \ref{c-uncert} and Table \ref{b-uncert}, respectively. 

\begin{table}[!htb]
\caption{The cross sections (pb) for $B_c^{**}$ production by taking different value of
$m_c$ and fixing $m_b = 4.90~\rm{GeV}$ at the $\sqrt{S} = 1.30 ~\rm{TeV}$ LHeC.}
\begin{center}
\begin{tabular}{cccc }
\hline
$m_c ~(\rm{GeV}) $ & 1.30 & 1.50 & 1.70 \\ \hline
$\sigma_{\gamma c}(^1P_1) $ & $6.33\times10^{-2}$ & $4.87\times10^{-2}$ & $3.80\times10^{-2}$ \\
$\sigma_{\gamma c}(^3P_0) $ & $1.52\times10^{-2}$ & $1.28\times10^{-2}$ & $1.08\times10^{-2}$ \\
$\sigma_{\gamma c}(^3P_1) $ & $3.44\times10^{-2}$ & $2.68\times10^{-2}$ & $2.12\times10^{-2}$ \\
$\sigma_{\gamma c}(^3P_2) $ & $9.58\times10^{-2}$ & $7.13\times10^{-2}$ & $5.42\times10^{-2}$ \\
$\sigma_{\gamma \bar{b}}(^1P_1) $ & 5.63 & 2.30 & 1.05 \\
$\sigma_{\gamma \bar{b}}(^3P_0) $ & 2.52 & 1.10 & $5.31\times10^{-1}$ \\
$\sigma_{\gamma \bar{b}}(^3P_1) $ & 5.48 & 2.27 & 1.05 \\
$\sigma_{\gamma \bar{b}}(^3P_2) $ & 12.5 & 5.22 & 2.43 \\
$\sigma_{\gamma g}(^1P_1) $ & $2.48\times10^{-1}$ & $1.27\times10^{-1}$ & $7.13\times10^{-2}$ \\
$\sigma_{\gamma g}(^3P_0) $ & $5.49\times10^{-2}$ & $3.08\times10^{-2}$ & $1.88\times10^{-2}$ \\
$\sigma_{\gamma g}(^3P_1) $ & $1.55\times10^{-1}$ & $7.72\times10^{-2}$ & $4.22\times10^{-2}$ \\
$\sigma_{\gamma g}(^3P_2) $ & $5.93\times10^{-1}$ & $2.99\times10^{-1}$ & $1.65\times10^{-1}$ \\
Total & 27.4 & 11.6 & 5.48 \\
\hline
\end{tabular}
\label{c-uncert}
\end{center}
\end{table}

\begin{table}[!htb]
\caption{The cross sections (pb) for $B_c^{**}$ production by taking different value
of $m_b$ and fixing $m_c = 1.50~\rm{GeV}$ at the $\sqrt{S} = 1.30 ~\rm{TeV}$ LHeC.}
\begin{center}
\begin{tabular}{cccc }
\hline
$m_b ~(\rm{GeV}) $ & 4.50 & 4.90 & 5.30 \\ \hline
$\sigma_{\gamma c}(^1P_1) $ & $7.39\times10^{-2}$ & $4.87\times10^{-2}$ & $3.33\times10^{-2}$ \\
$\sigma_{\gamma c}(^3P_0) $ & $2.05\times10^{-2}$ & $1.28\times10^{-2}$ & $8.29\times10^{-3}$ \\
$\sigma_{\gamma c}(^3P_1) $ & $4.12\times10^{-2}$ & $2.68\times10^{-2}$ & $1.81\times10^{-2}$ \\
$\sigma_{\gamma c}(^3P_2) $ & $1.07\times10^{-1}$ & $7.13\times10^{-2}$ & $4.95\times10^{-2}$ \\
$\sigma_{\gamma \bar{b}}(^1P_1) $ & 1.08 & 2.30 & 3.13 \\
$\sigma_{\gamma \bar{b}}(^3P_0) $ & $4.93\times10^{-1}$ & 1.10 & 1.47 \\
$\sigma_{\gamma \bar{b}}(^3P_1) $ & 1.03 & 2.27 & 3.10 \\
$\sigma_{\gamma \bar{b}}(^3P_2) $ & 2.46 & 5.22 & 7.04 \\
$\sigma_{\gamma g}(^1P_1) $ & $1.77\times10^{-1}$ & $1.27\times10^{-1}$ & $9.39\times10^{-2}$ \\
$\sigma_{\gamma g}(^3P_0) $ & $4.54\times10^{-2}$ & $3.08\times10^{-2}$ & $2.17\times10^{-2}$ \\
$\sigma_{\gamma g}(^3P_1) $ & $1.06\times10^{-1}$ & $7.72\times10^{-2}$ & $5.78\times10^{-2}$ \\
$\sigma_{\gamma g}(^3P_2) $ & $4.10\times10^{-1}$ & $2.99\times10^{-1}$ & $2.23\times10^{-1}$ \\
Total & 6.05 & 11.6 & 15.2 \\
\hline
\end{tabular}
\label{b-uncert}
\end{center}
\end{table}

\par
From the tables we can see that for
most of the channels, the cross sections decrease when the $c$ or $b$-quark mass becomes larger.
The only exception is for the $\gamma + \bar{b}$ channel, whose cross sections increase when the
$b$-quark mass becomes larger. The cross sections are much more sensitive to $m_c$ than to $m_b$, however, the $\gamma + \bar{b}$ channel provides most of the contributions, thus we conclude that the total cross section decreases when the $c$-quark mass
becomes larger or the $b$-quark mass becomes smaller. Besides, it is clear that the cross sections of P-wave states are more
sensitive to quark masses than the S-wave states as declared in \cite{Yang:2010yg,prod22}.
By summing up the cross sections of all production channels and their mass uncertainties,
we obtain the total cross sections as
\begin{eqnarray}
\sigma^{\rm{Total}}_{\rm{LHeC}} &=& 11.6^{+15.9}_{-6.10} \, {\rm{pb}}, ~{{\rm{for}}~m_c=1.50\pm0.20~{\rm{GeV}}}, \\
\sigma^{\rm{Total}}_{\rm{LHeC}} &=& 11.6^{+3.67}_{-5.53} \, {\rm{pb}}, ~{{\rm{for}}~m_b=4.90\pm0.40~{\rm{GeV}}},
\end{eqnarray}
and by adding the errors from two mass uncertainties in quadrature, we finally obtain
\begin{eqnarray}
\sigma^{\rm{Total}}_{\rm{LHeC}} &=& 11.6^{+16.3}_{-8.24} \, {\rm{pb}},
~{{\rm{for}}~m_b=4.90\pm0.40~{\rm{GeV}} ~{\rm{and}} ~m_c=1.50\pm0.20~{\rm{GeV}}}.
\end{eqnarray}
That means the contribution from total cross section of the P-wave $B_c^{**}$ states can be about $11\%\sim27\%$
of that from the S-wave $B_c^{(*)}$ state production ($59.0^{+46.2}_{-28.3}$ pb)\cite{ep-S} in the range of the uncertainties on heavy quark
masses. So these excited state contributions should be taken into consideration, especially for
the future high energy and high luminosity colliders.

\par
\begin{figure}[htbp]
\includegraphics[scale=0.55]{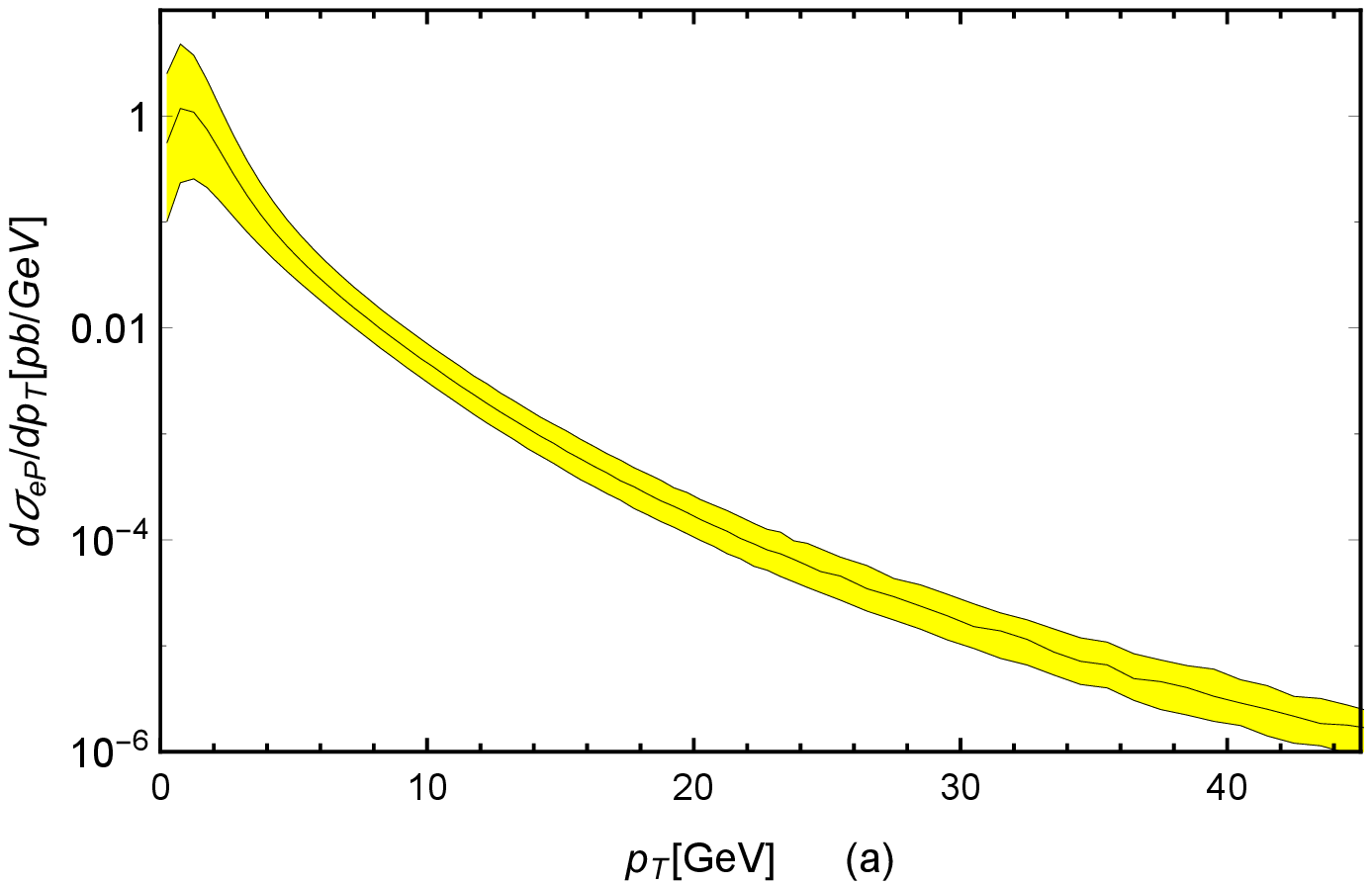}
\hspace{0in}
\includegraphics[scale=0.55]{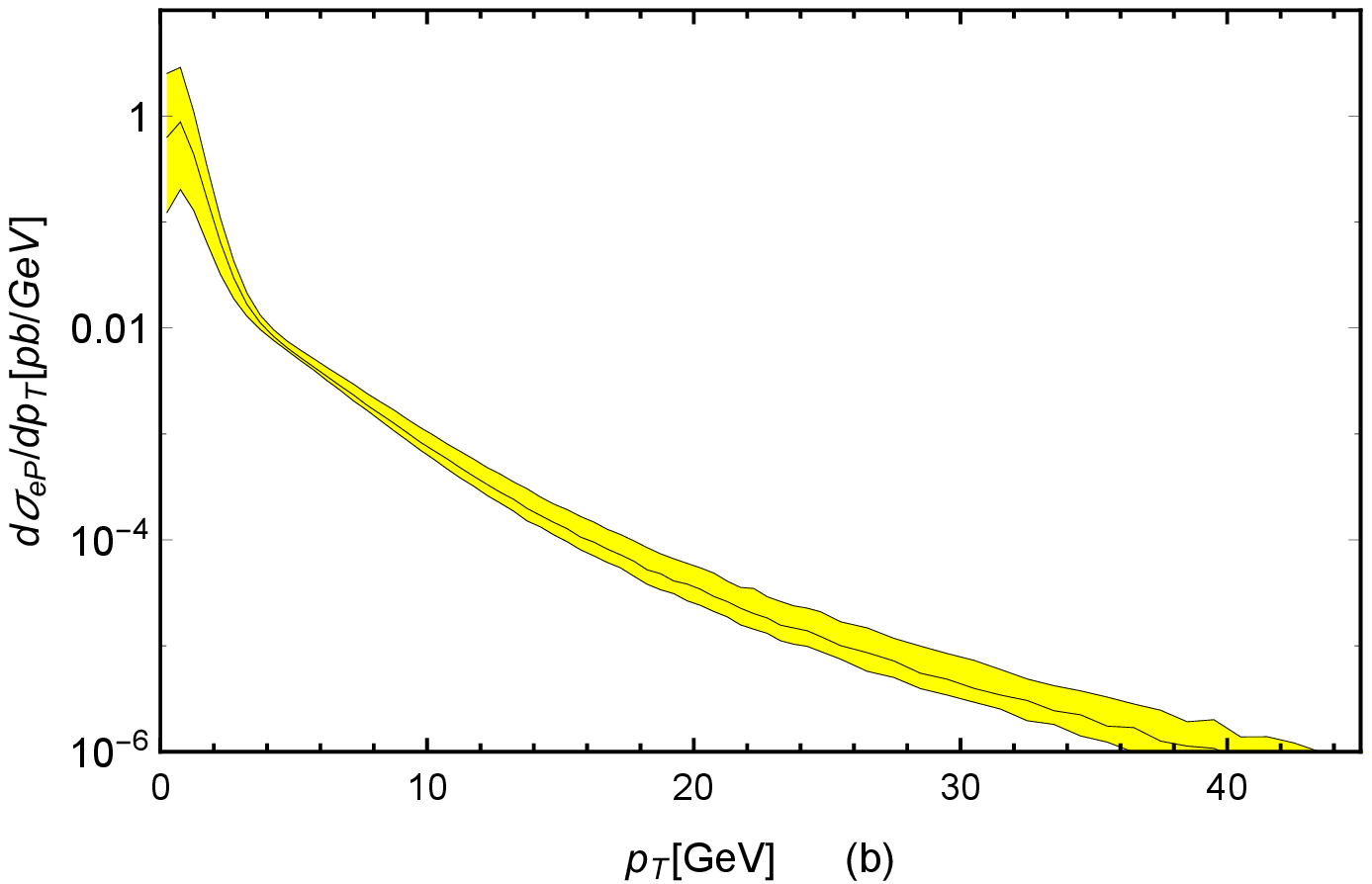}
\hspace{0in}
\vfill
\includegraphics[scale=0.55]{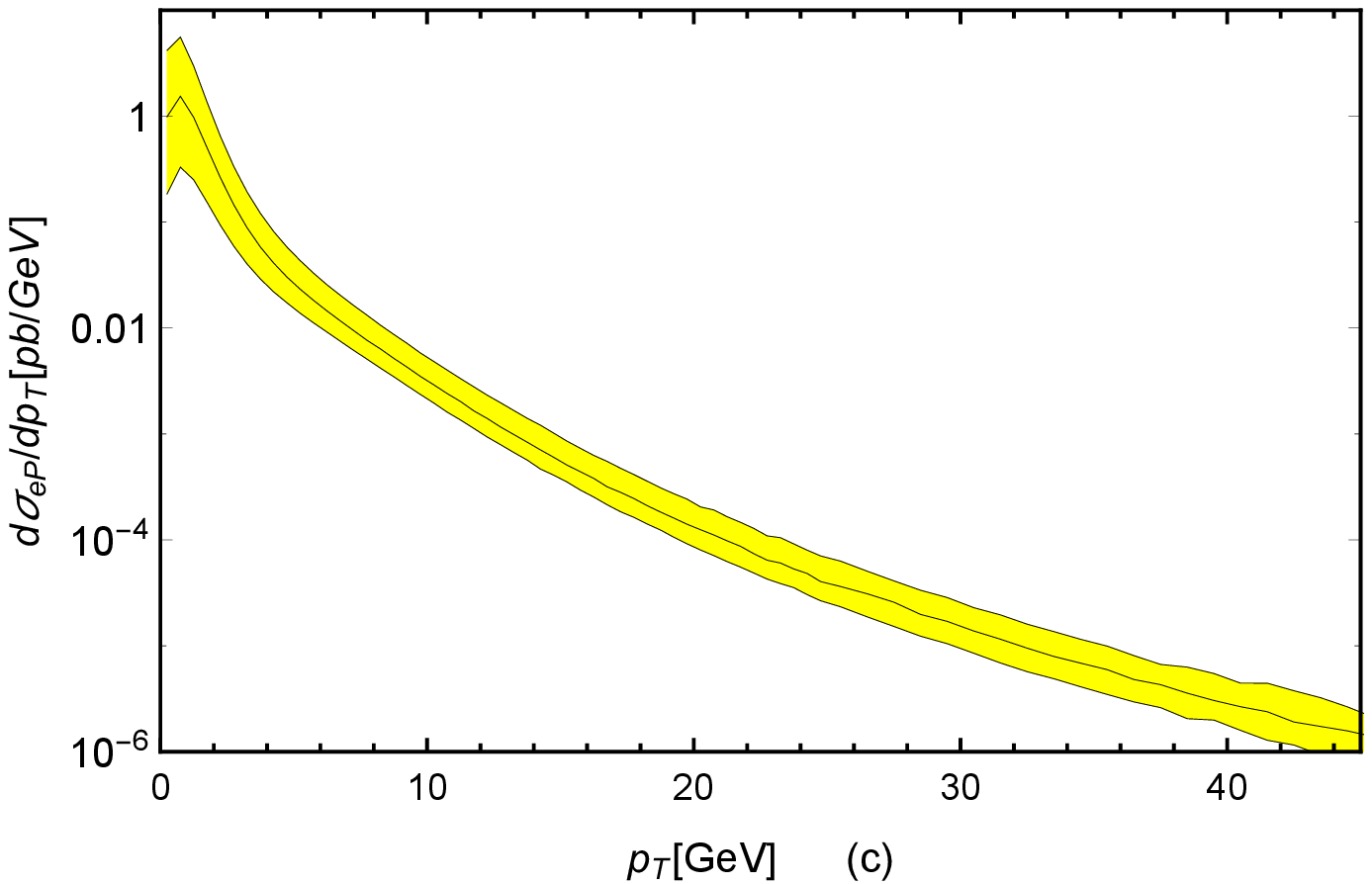}
\hspace{0in}
\includegraphics[scale=0.55]{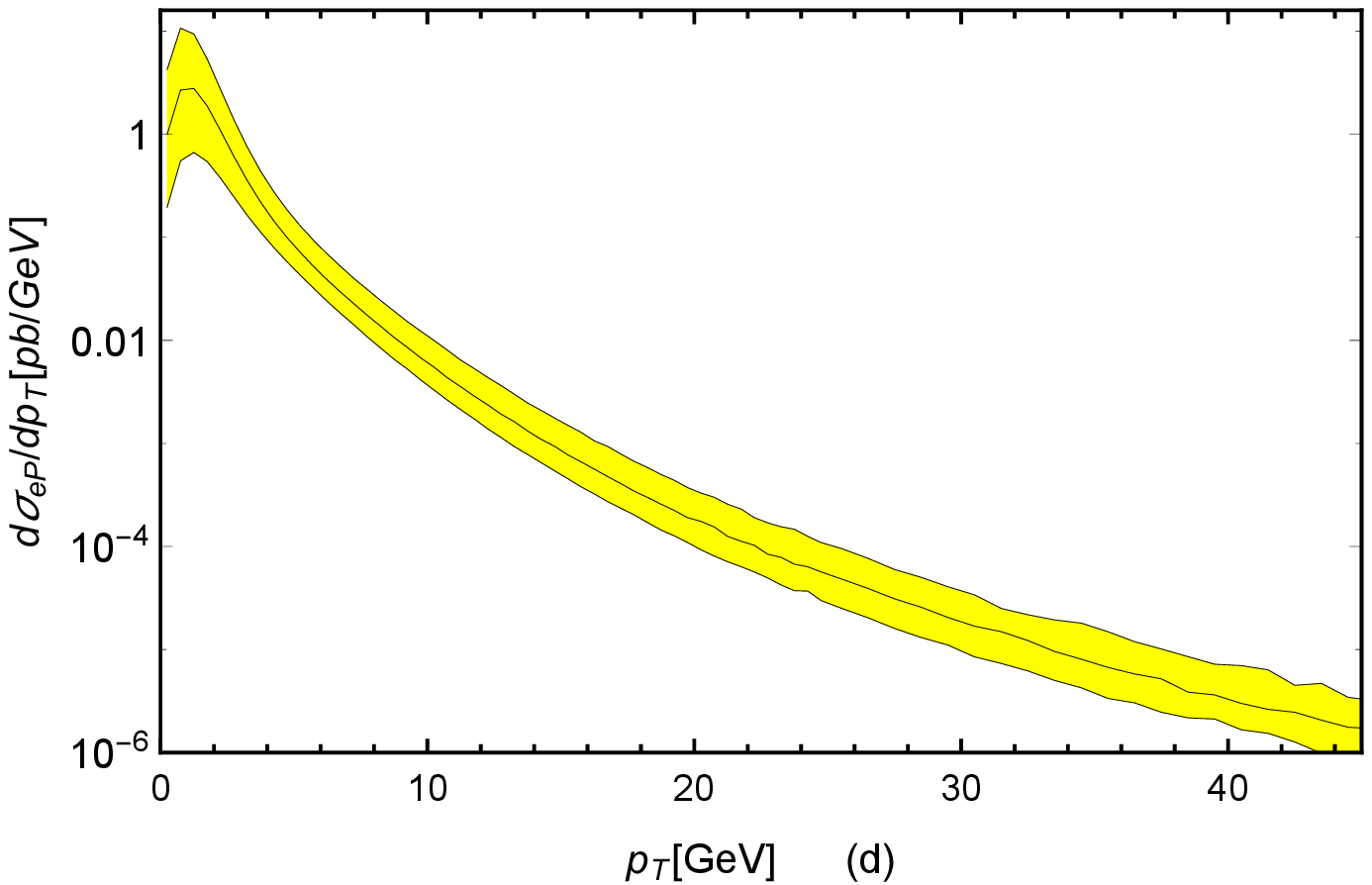}
\hspace{0in}
\caption{Uncertainties of the transverse momentum distributions from the quark masses for the $B_c^{**}$ meson photoproduction at the
$\sqrt{S} = 1.30 ~\rm{TeV}$ LHeC, where (a), (b), (c) and (d) denotes $^1P_1$, $^3P_0$, $^3P_1$ and $^3P_2$ $B_c^{**}$ state, respectively.}
\label{ptdist}
\end{figure}
\begin{figure}[htbp]
\includegraphics[scale=0.55]{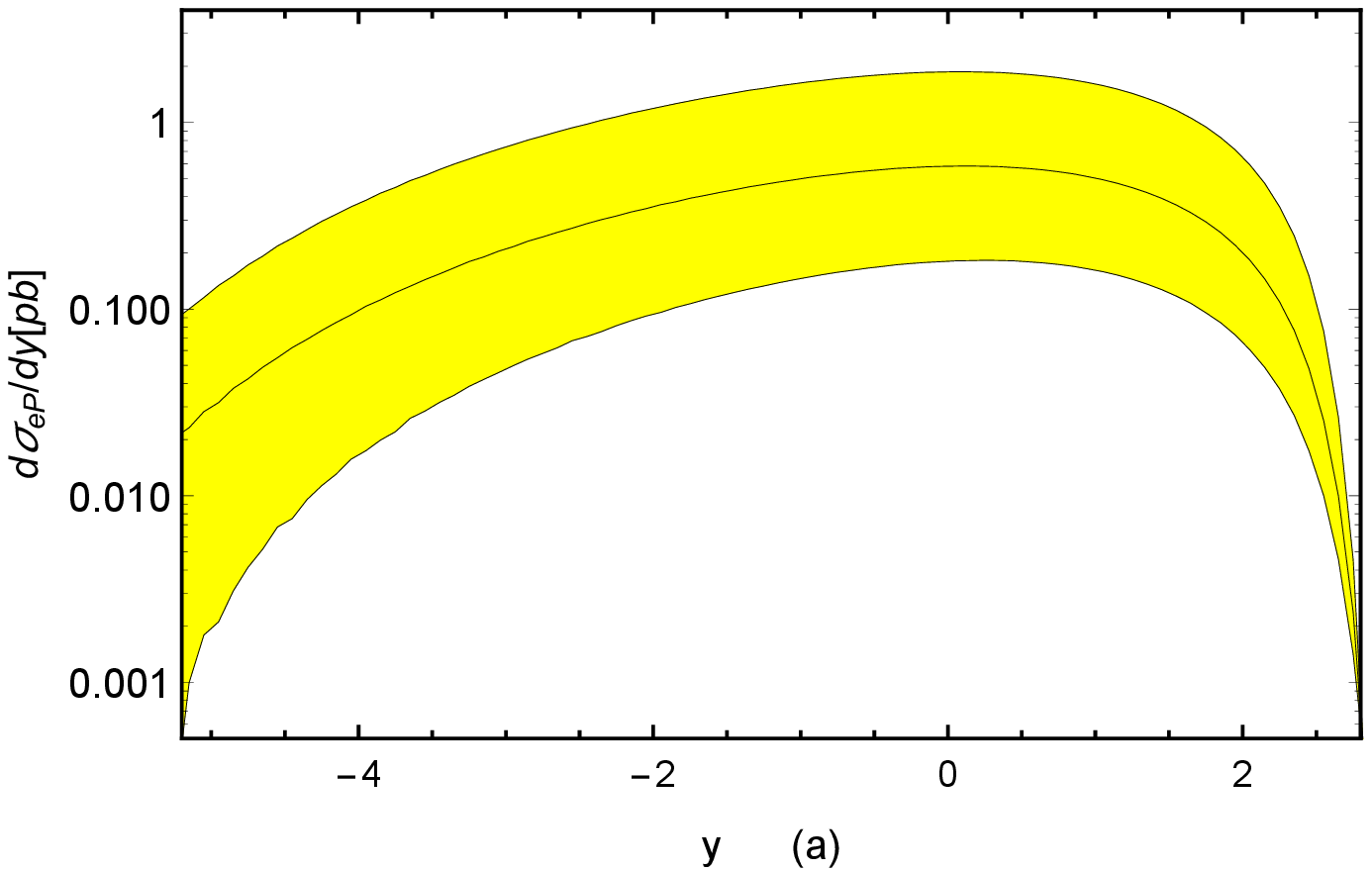}
\hspace{0in}
\includegraphics[scale=0.55]{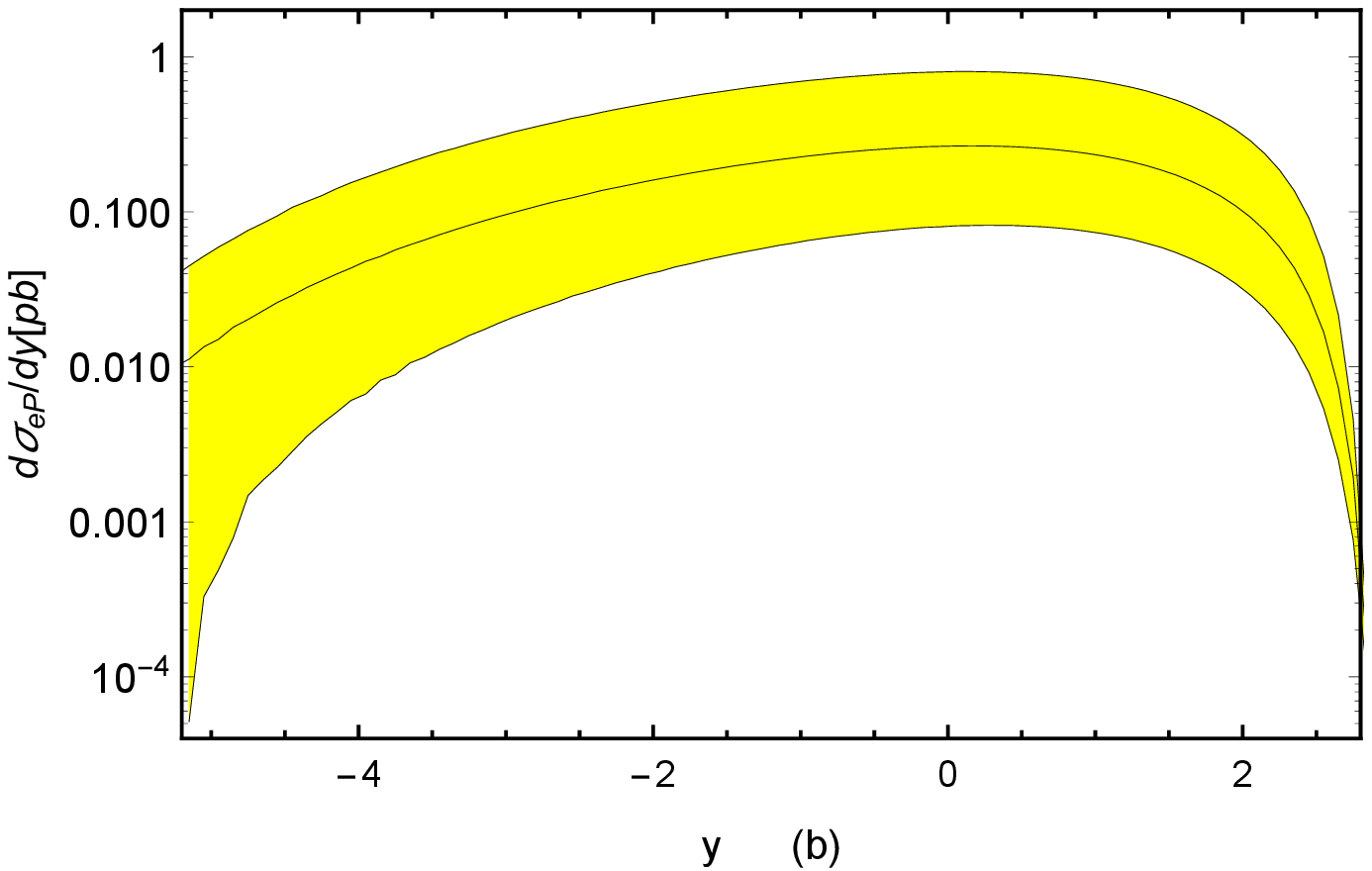}
\hspace{0in}
\vfill
\includegraphics[scale=0.55]{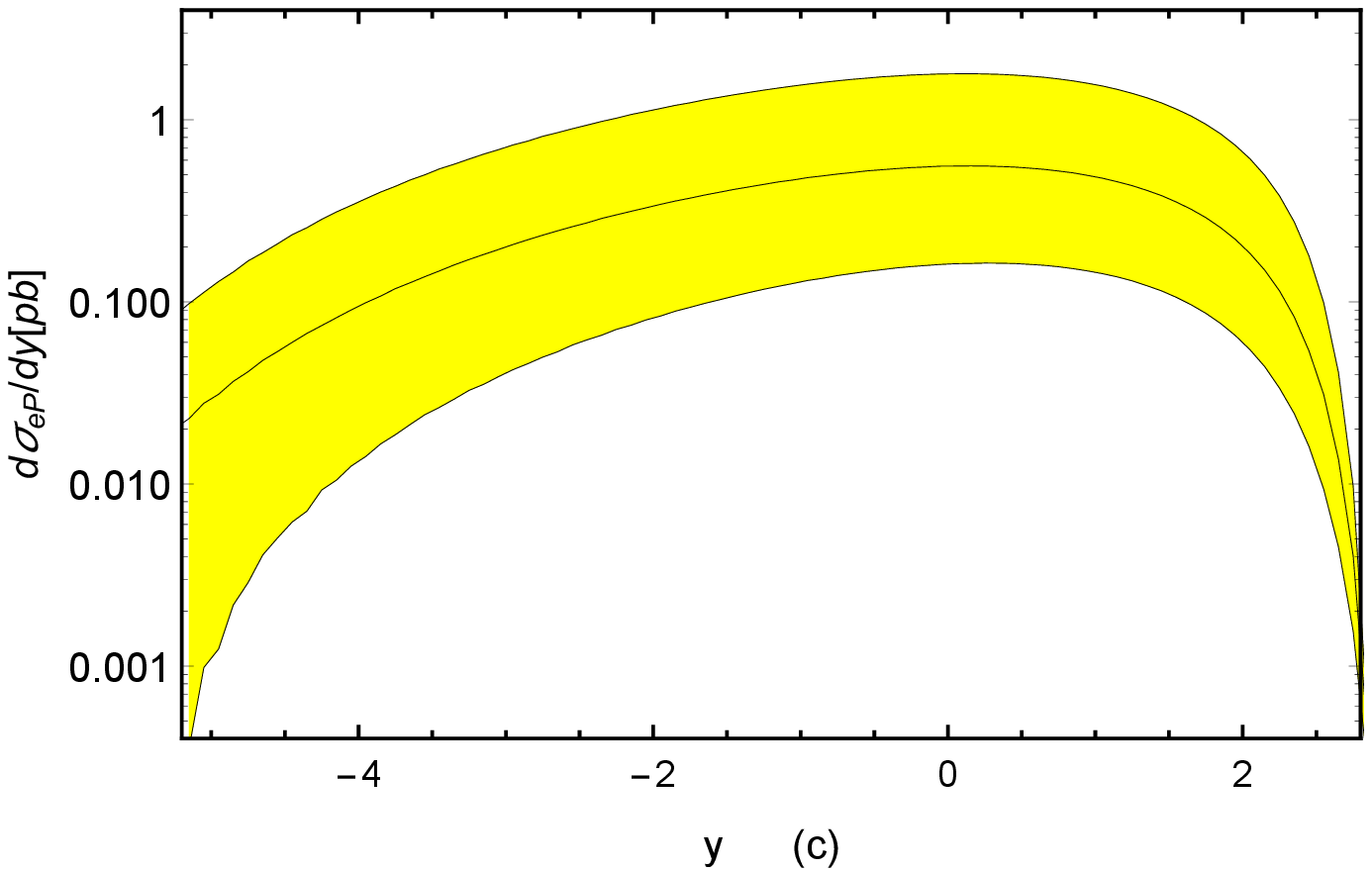}
\hspace{0in}
\includegraphics[scale=0.55]{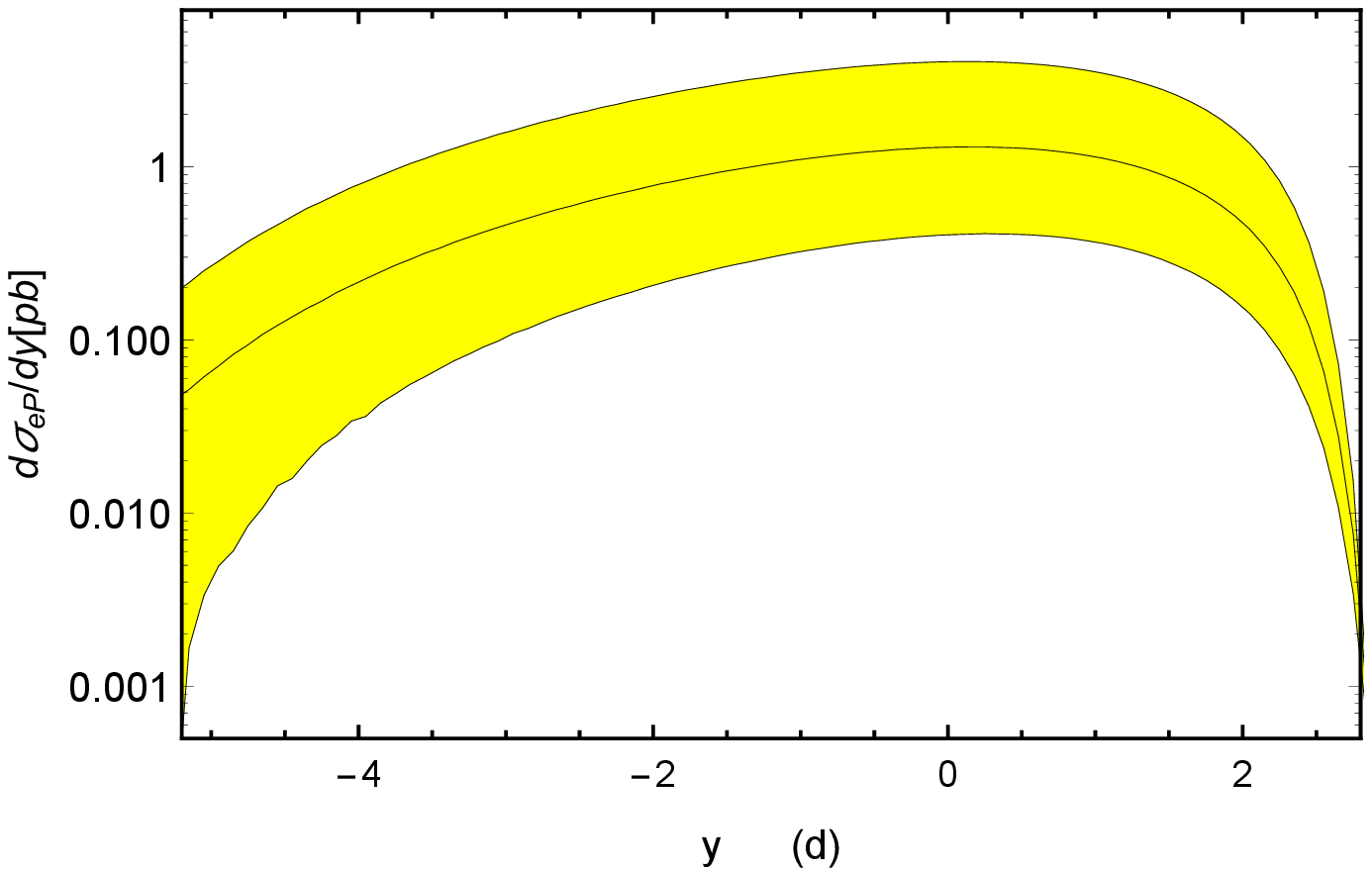}
\hspace{0in}
\caption{Uncertainties of the rapidity distributions from the quark masses for the $B_c^{**}$ meson photoproduction at the
$\sqrt{S} = 1.30 ~\rm{TeV}$ LHeC, where (a), (b), (c) and (d) denotes $^1P_1$, $^3P_0$, $^3P_1$ and $^3P_2$ $B_c^{**}$ state, respectively.}
\label{ydist}
\end{figure}
\begin{figure}[htbp]
\includegraphics[scale=0.55]{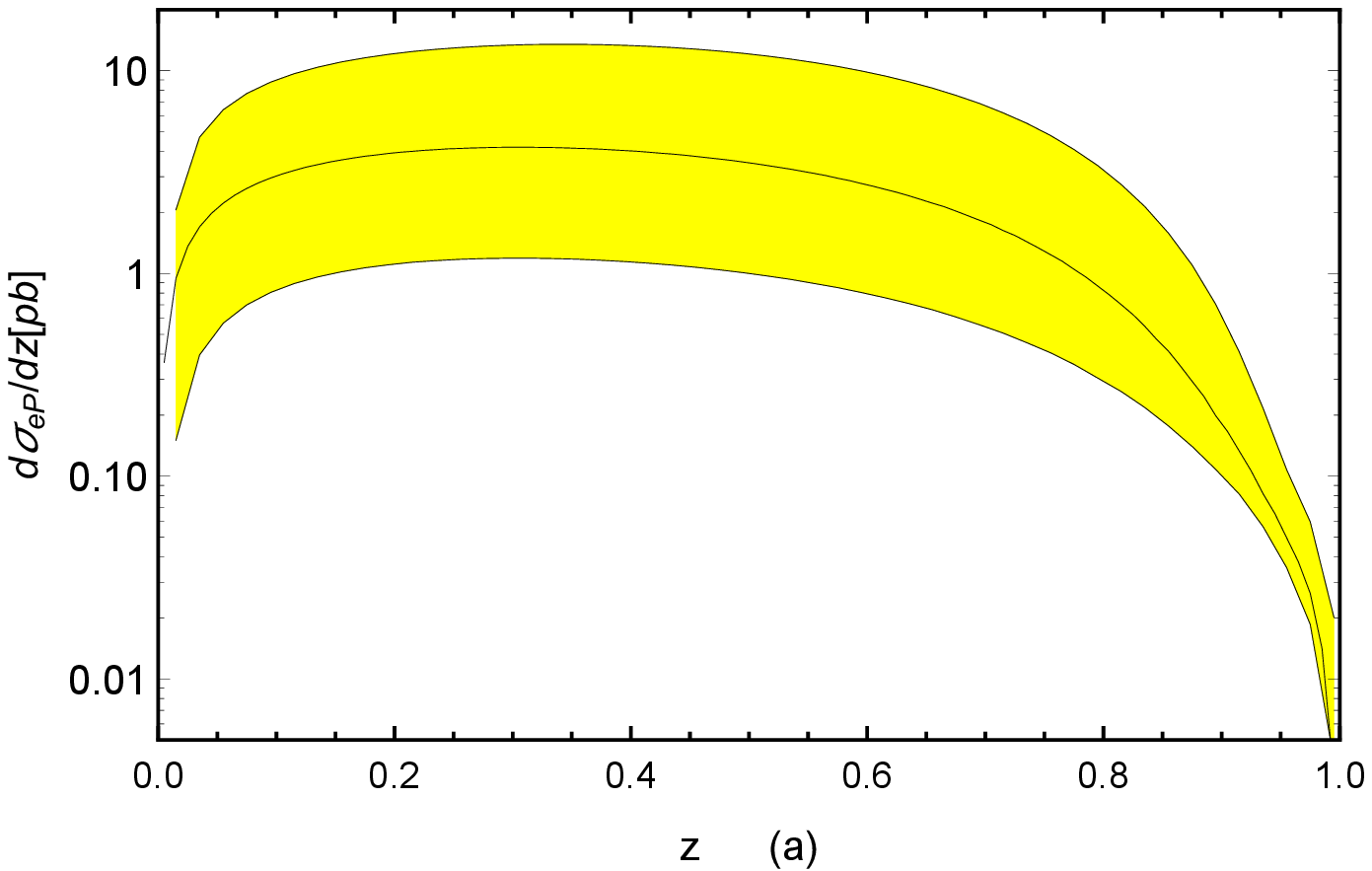}
\hspace{0in}
\includegraphics[scale=0.55]{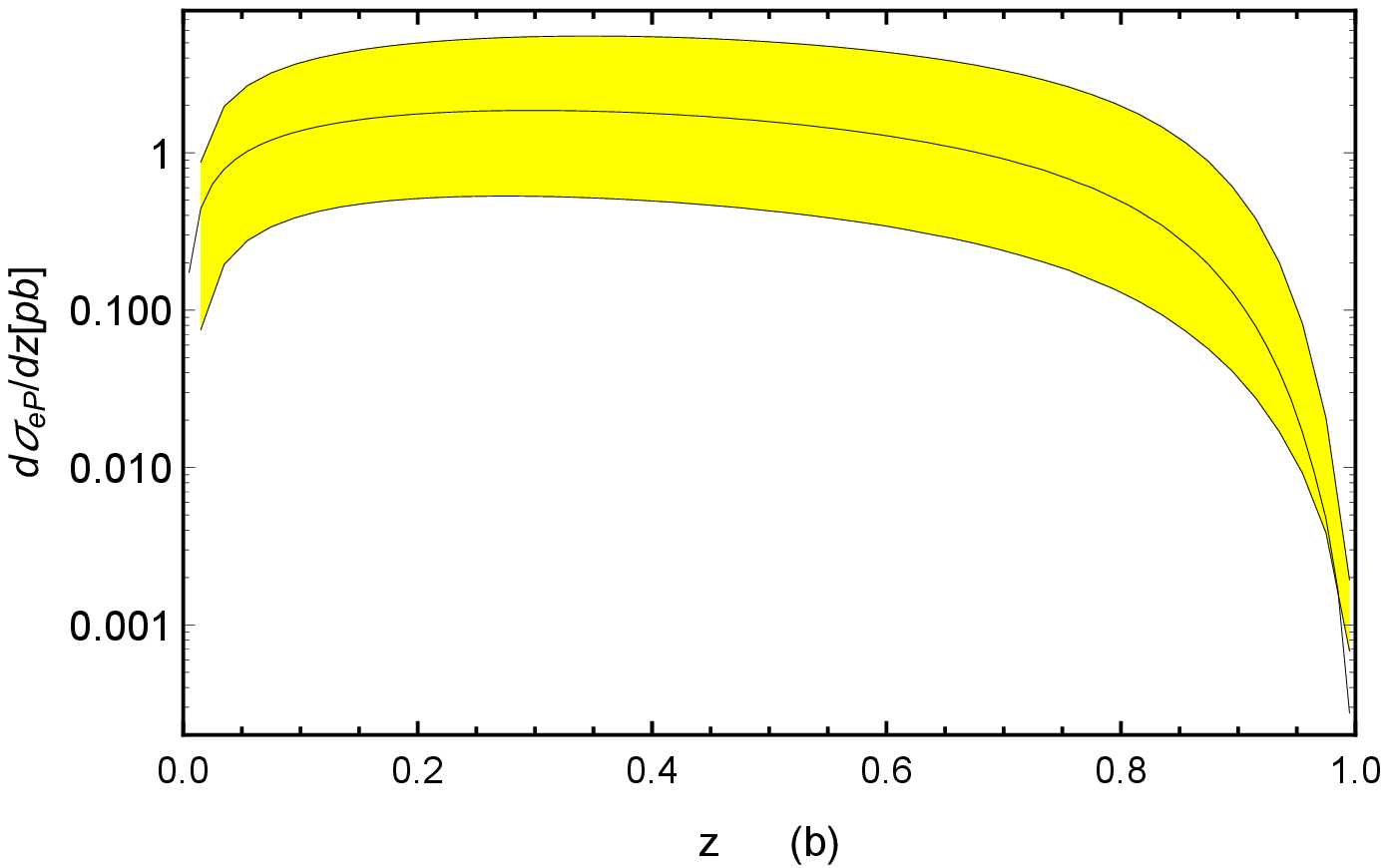}
\hspace{0in}
\vfill
\includegraphics[scale=0.55]{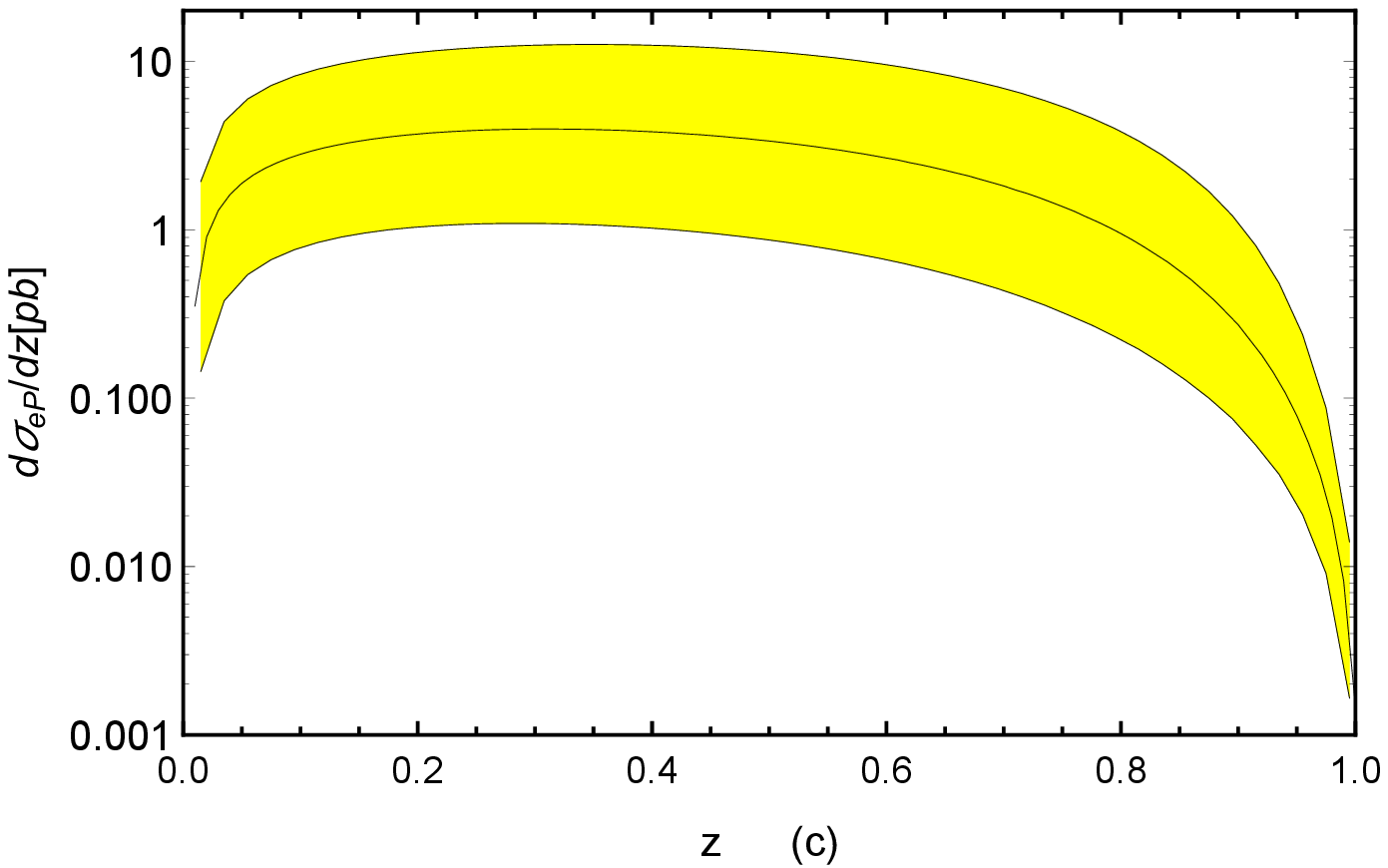}
\hspace{0in}
\includegraphics[scale=0.55]{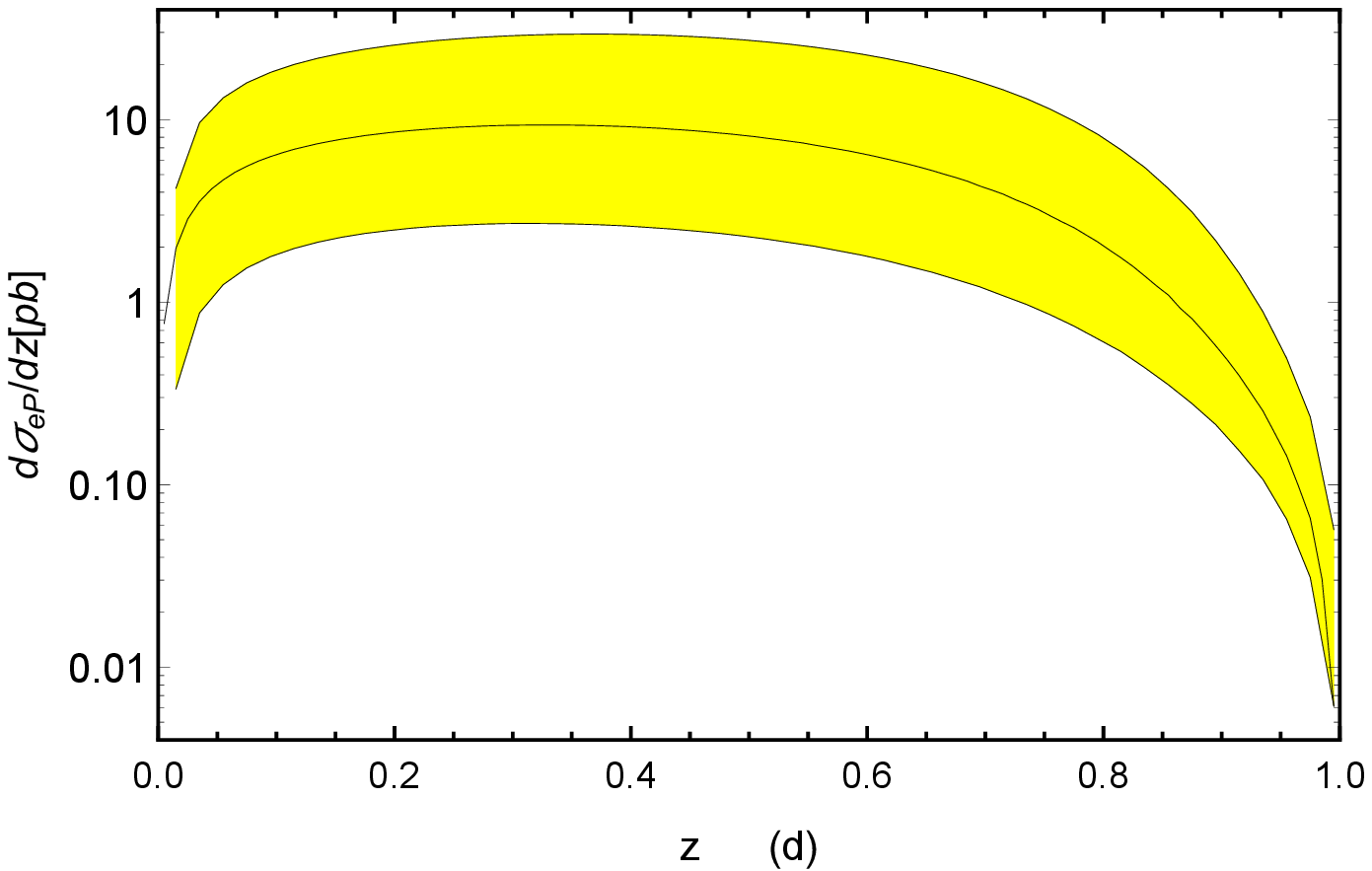}
\hspace{0in}
\caption{Uncertainties of the differential cross sections $d\sigma/dz$ versus $z$ from the quark masses for the $B_c^{**}$ meson photoproduction at the
$\sqrt{S} = 1.30 ~\rm{TeV}$ LHeC, where (a), (b), (c) and (d) denotes $^1P_1$, $^3P_0$, $^3P_1$ and $^3P_2$ $B_c^{**}$ state, respectively.}
\label{zdist}
\end{figure}

With all the channels summed up, the shaded bands in Figs. \ref{ptdist}, \ref{ydist}, \ref{zdist} show the various uncertainties associated to the quark masses clearly.
The central values correspond to $m_c=1.50$GeV and $m_b=4.90$GeV, while the upper and lower bounds are obtained by setting $m_c=1.30$GeV, $m_b=5.30$GeV and
$m_c=1.70$GeV, $m_b=4.50$GeV, respectively. 


\section{SUMMARY}
\par
In this paper, we studied the P-wave excited $B_c^{**}$ ($^1P_1$ and $^3P_J$ with $J=0,1,2$) meson
photoproduction at the LHeC and three photoproduction channels, i.e.,
$e^-+P \to \gamma +g \to B_c^{**}+b+\bar{c}$, $e^-+P \to \gamma +c \to B_c^{**}+b$
and $e^-+P \to \gamma +\bar{b} \to B_c^{**}+\bar{c}$  are considered here.
It is found that the production of the P-wave $B_c^{**}$ states can contribute about $20\%$
of the S-wave $B_c^{(*)}$ production at the LHeC and FCC-$ep$, if considering the fact that
almost all of the P-wave $B_c^{**}$ states decay to the ground state $B_c(^1S_0)$. Therefore, for
the studying the production of $B_c$ meson, the P-wave excited states should also be included.
Taking the most prominent errors from the heavy quark masses, $m_b=4.90\pm0.40$ GeV and
$m_c=1.50\pm0.20$ GeV, into account, we would expect to accumulate about
$(2.48^{+3.55}_{-1.75}) \times 10^4$ $B_c^{**}({^{1}P_1})$, $(1.14^{+1.49}_{-0.82})
\times 10^4$  $B_c^{**}({^{3}P_0})$, $(2.38^{+3.39}_{-1.74}) \times 10^4$  $B_c^{**}({^{3}P_1})$
and $(5.59^{+7.84}_{-3.93}) \times 10^4$  $B_c^{**}({^{3}P_2})$ events at the LHeC in one operation
year with $\sqrt{S}=1.30$ TeV collision energy and the luminosity ${\cal L}= 10^{33}$ cm$^{-2}$s$^{-1}$.
We find that the dominant contributions come from low $p_T$ region of  $\gamma +\bar{b}$
channel, so it is possible to directly measure
the P-wave $B_c^{**}$ states at the LHeC and FCC-$ep$ by using low $p_T$ tagging technology, and thus is helpful to understand
the mass spectrum of ($c\bar{b}$) bound states and to test the potential models.

\vskip 5mm
\par
\noindent{\large\bf Acknowledgments:}
This work is supported in part by the National Natural Science Foundation of China
(No.11775211, No.11405173, No.11535002) and the CAS Center for Excellence in Particle Physics (CCEPP).



\section*{Appendix A: Feynman diagrams and hard-scattering amplitudes}

\par
\begin{figure}[htb]
\includegraphics[scale=1]{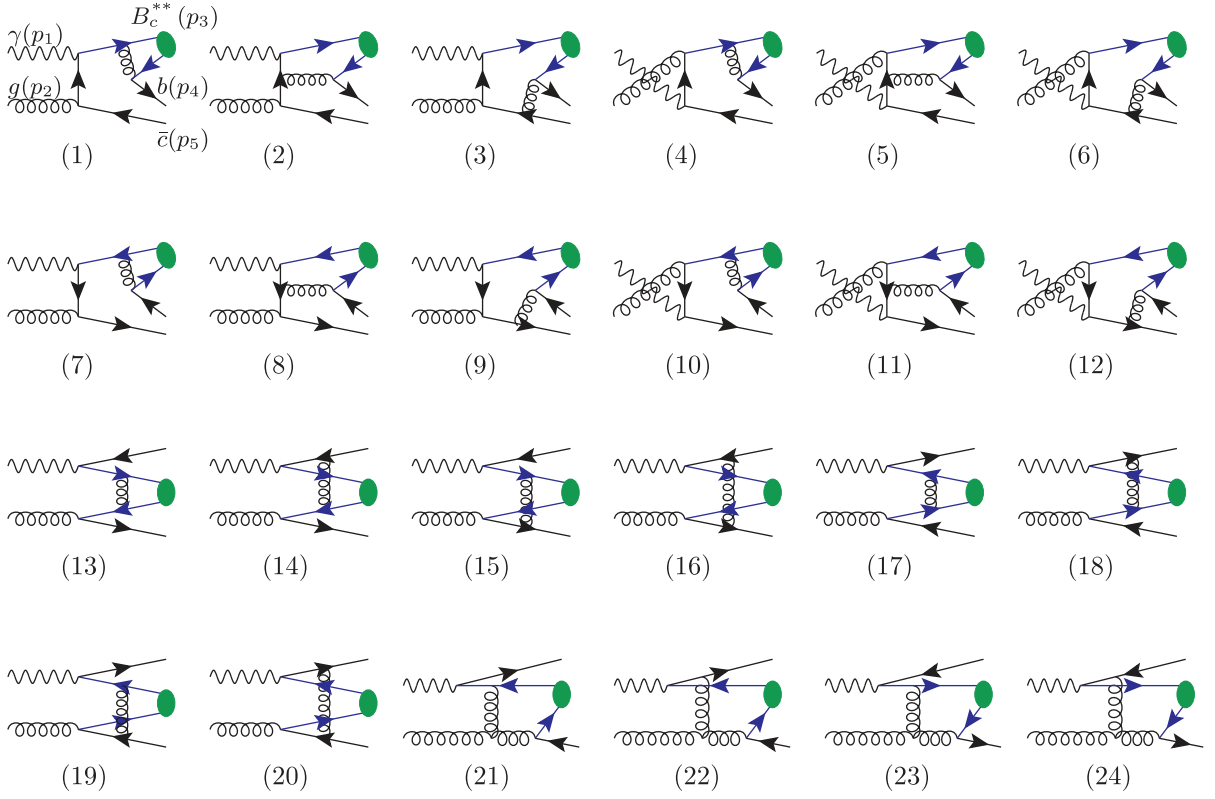}
\centering
\hspace{0in}
\caption{Feynman diagrams for the subprocess $\gamma  + g  \to B_c^{**}   + b  + \bar{c} $.}
\label{figfd1}
\end{figure}
\begin{figure}[htb]
\includegraphics[scale=0.75]{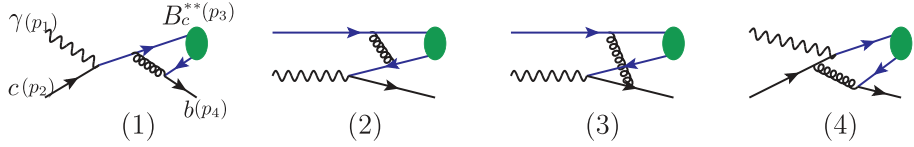}
\centering
\hspace{0in}
\caption{Feynman diagrams for the subprocess $\gamma  + c  \to B_c^{**}   + b $. The Feynman diagrams for the subprocess $\gamma  + \bar{b}  \to B_c^{**}   + \bar{c} $ can be obtained by the replacements $c \to \bar{b}$ and $b \to \bar{c}$.} \label{figfd2}
\end{figure}

We present the Feynman diagrams for $\gamma+g\to B_c^{**}+\bar{c}+b$ and $\gamma+c/\bar{b}\to B_c^{**}+b/\bar{c}$ in Figs.(\ref{figfd1}, \ref{figfd2}). There are twenty-four hard-scattering amplitudes $T_j$ ($T={ \sum_{j=1}^{24} }{T_j}$) for the subprocess $\gamma(p_1)+g(p_2)\to B_c^{**}(p_3)+\bar{c}(p_4)+b(p_5)$ which can be expressed as

\begin{eqnarray}
i T_{1} &=& ieg^3Q_c \mathcal{C}_1 \bar{u}_{s^{\prime}}(p_4) \gamma^{\mu} \frac{\Pi(p_3)}{(p_{32} + p_4)^2} \gamma_{\mu} \frac{ \not\! p_3 + \not\! p_4 + m_c }{(p_3 + p_4)^2 - m_c^2} \not\! \epsilon (p_1) \frac{\not\! p_2 - \not\! p_5 + m_c}{(p_2 - p_5)^2 - m_c^2} \not\! \epsilon(p_2) v_{s}(p_5),
 \nonumber \\
i T_{2} &=& i eg^3Q_c\mathcal{C}_2 \bar{u}_{s^{\prime}}(p_4) \gamma^{\mu} \frac{\Pi(p_3)}{(p_{32} + p_4)^2} \not\! \epsilon (p_1) \frac{\not\!p_{31} - \not\! p_1 + m_c }{(p_1 - p_{31})^2 - m_c^2} \gamma_{\mu} \frac{\not\!p_2 - \not\!p_5 + m_c^2}{(p_2 - p_5)^2 - m_c^2} \not\! \epsilon(p_2) v_{s}(p_5),
\nonumber \\
i T_{3} &=& ieg^3Q_c \mathcal{C}_3 \bar{u}_{s^{\prime}}(p_4) \gamma^{\mu} \frac{\Pi(p_3)}{(p_{32} + p_4)^2} \not\! \epsilon (p_1) \frac{\not\!p_{31} - \not\! p_1 + m_c }{(p_1 - p_{31})^2 - m_c^2} \not\! \epsilon(p_2) \frac{-\not\!p_{32} - \not\!p_4 - \not\!p_5 + m_c}{(p_{32} + p_4 + p_5)^2 - m_c^2} \gamma_{\mu} v_{s}(p_5),
\nonumber \\
i T_{4} &=& ieg^3Q_c \mathcal{C}_4 \bar{u}_{s^{\prime}}(p_4) \gamma^{\mu} \frac{\Pi(p_3)}{(p_{32} + p_4)^2} \gamma_{\mu} \frac{ \not\! p_3 + \not\! p_4 + m_c }{(p_3 + p_4)^2 - m_c^2} \not\! \epsilon(p_2) \frac{\not\! p_1 - \not\! p_5 + m_c}{(p_1 - p_5)^2 - m_c^2} \not\! \epsilon (p_1) v_{s}(p_5),
\nonumber \\
i T_{5} &=& ieg^3Q_c \mathcal{C}_5 \bar{u}_{s^{\prime}}(p_4) \gamma^{\mu} \frac{\Pi(p_3)}{(p_{32} + p_4)^2} \not\! \epsilon (p_2) \frac{\not\!p_{31} - \not\! p_2 + m_c }{(p_2 - p_{31})^2 - m_c^2} \gamma_{\mu} \frac{\not\!p_1 - \not\!p_5 + m_c^2}{(p_1 - p_5)^2 - m_c^2} \not\! \epsilon(p_1) v_{s}(p_5),
\nonumber \\
i T_{6} &=& ieg^3 Q_c\mathcal{C}_6 \bar{u}_{s^{\prime}}(p_4) \gamma^{\mu} \frac{\Pi(p_3)}{(p_{32} + p_4)^2} \not\! \epsilon (p_2) \frac{\not\!p_{31} - \not\! p_2 + m_c }{(p_2 - p_{31})^2 - m_c^2} \not\! \epsilon(p_1) \frac{-\not\!p_{32} - \not\!p_4 - \not\!p_5 + m_c}{(p_{32} + p_4 + p_5)^2 - m_c^2} \gamma_{\mu} v_{s}(p_5),\nonumber 
\end{eqnarray}
\begin{eqnarray}
i T_{7} &=& i eg^3Q_b\mathcal{C}_7 \bar{u}_{s^{\prime}}(p_4) \not\! \epsilon (p_2) \frac{\not\!p_4 - \not\!p_2 + m_b}{(p_4 - p_2)^2 - m_b^2} \not\! \epsilon (p_1) \frac{-\not\!p_3 - \not\!p_5 + m_b}{(p_3 + p_5)^2 - m_b^2} \gamma^{\mu} \frac{\Pi(p_3)}{(p_{31} + p_5)^2} \gamma_{\mu} v_{s}(p_5),
\nonumber \\
i T_{8} &=& eg^3Q_b \mathcal{C}_8 \bar{u}_{s^{\prime}}(p_4) \not\! \epsilon (p_2) \frac{\not\!p_4 - \not\!p_2 + m_b}{(p_4 - p_2)^2 - m_b^2} \gamma^{\mu} \frac{\not\!p_1 - \not\!p_{32} + m_b}{(p_1 - p_{32})^2 - m_b^2} \not\! \epsilon (p_1) \frac{\Pi(p_3)}{(p_{31} + p_5)^2} \gamma_{\mu} v_{s}(p_5),
\nonumber \\
i T_{9} &=& ieg^3Q_b \mathcal{C}_9 \bar{u}_{s^{\prime}}(p_4) \gamma^{\mu} \frac{\not\!p_{31} + \not\!p_4 + \not\!p_5 + m_b}{(p_{31} + p_4 + p_5)^2 - m_b^2} \not\! \epsilon (p_2) \frac{\not\!p_1 - \not\!p_{32} + m_b}{(p_1 - p_{32})^2 - m_b^2} \not\! \epsilon (p_1) \frac{\Pi(p_3)}{(p_{31} + p_5)^2} \gamma_{\mu} v_{s}(p_5),
\nonumber \\
i T_{10} &=& ieg^3Q_b \mathcal{C}_{10} \bar{u}_{s^{\prime}}(p_4) \not\! \epsilon (p_1) \frac{\not\!p_4 - \not\!p_1 + m_b}{(p_4 - p_1)^2 - m_b^2} \not\! \epsilon (p_2) \frac{-\not\!p_3 - \not\!p_5 + m_b}{(p_3 + p_5)^2 - m_b^2} \gamma^{\mu} \frac{\Pi(p_3)}{(p_{31} + p_5)^2} \gamma_{\mu} v_{s}(p_5),
\nonumber \\
i T_{11} &=& ieg^3Q_b \mathcal{C}_{11} \bar{u}_{s^{\prime}}(p_4) \not\! \epsilon (p_1) \frac{\not\!p_4 - \not\!p_1 + m_b}{(p_4 - p_1)^2 - m_b^2} \gamma^{\mu} \frac{\not\!p_2 - \not\!p_{32} + m_b}{(p_2 - p_{32})^2 - m_b^2} \not\! \epsilon (p_2) \frac{\Pi(p_3)}{(p_{31} + p_5)^2} \gamma_{\mu} v_{s}(p_5),
\nonumber \\
i T_{12} &=& ieg^3Q_b \mathcal{C}_{12} \bar{u}_{s^{\prime}}(p_4) \gamma^{\mu} \frac{\not\!p_{31} + \not\!p_4 + \not\!p_5 + m_b}{(p_{31} + p_4 + p_5)^2 - m_b^2} \not\! \epsilon (p_1) \frac{\not\!p_2 - \not\!p_{32} + m_b}{(p_2 - p_{32})^2 - m_b^2} \not\! \epsilon (p_2) \frac{\Pi(p_3)}{(p_{31} + p_5)^2} \gamma_{\mu} v_{s}(p_5),\nonumber 
\end{eqnarray}
\begin{eqnarray}
i T_{13} &=& ieg^3Q_c \mathcal{C}_{13} \bar{u}_{s^{\prime}}(p_4) \not\! \epsilon (p_2) \frac{\not\!p_4 - \not\!p_2 + m_b^2}{(p_2 - p_4)^2 - m_b^2} \gamma^{\mu} \frac{\Pi(p_3)}{(p_1 - p_5 - p_{31})^2} \gamma_{\mu} \frac{\not\!p_1 - \not\!p_5 + m_c}{(p_1 - p_5)^2 - m_c^2} \not\! \epsilon (p_1) v_{s}(p_5),
\nonumber \\
i T_{14} &=& ieg^3Q_c \mathcal{C}_{14} \bar{u}_{s^{\prime}}(p_4) \not\! \epsilon (p_2) \frac{\not\!p_4 - \not\!p_2 + m_b^2}{(p_2 - p_4)^2 - m_b^2} \gamma^{\mu} \frac{\Pi(p_3)}{(p_1 - p_5 - p_{31})^2} \not\! \epsilon (p_1) \frac{\not\!p_{31}- \not\!p_1 + m_c}{(p_1 - p_{31})^2 - m_c^2} \gamma_{\mu} v_{s}(p_5),
\nonumber \\
i T_{15} &=& ieg^3Q_c \mathcal{C}_{15} \bar{u}_{s^{\prime}}(p_4) \gamma^{\mu} \frac{\not\!p_2 - \not\!p_{32} + m_b}{(p_2 - p_{32})^2 - m_b^2} \not\! \epsilon (p_2) \frac{\Pi(p_3)}{(p_1 - p_5 - p_{31})^2} \gamma_{\mu} \frac{\not\!p_1 - \not\!p_5 + m_c}{(p_1 - p_5)^2 - m_c^2} \not\! \epsilon (p_1) v_{s}(p_5),
\nonumber \\
i T_{16} &=& ieg^3Q_c \mathcal{C}_{16} \bar{u}_{s^{\prime}}(p_4) \gamma^{\mu} \frac{\not\!p_2 - \not\!p_{32} + m_b}{(p_2 - p_{32})^2 - m_b^2} \not\! \epsilon (p_2) \frac{\Pi(p_3)}{(p_1 - p_5 - p_{31})^2} \not\! \epsilon (p_1) \frac{\not\!p_{31}- \not\!p_1 + m_c}{(p_1 - p_{31})^2 - m_c^2} \gamma_{\mu} v_{s}(p_5),\nonumber 
\end{eqnarray}
\begin{eqnarray}
i T_{17} &=& ieg^3Q_b \mathcal{C}_{17} \bar{u}_{s^{\prime}}(p_4) \not\! \epsilon (p_1) \frac{\not\!p_4 - \not\!p_1 + m_b^2}{(p_1 - p_4)^2 - m_b^2} \gamma^{\mu} \frac{\Pi(p_3)}{(p_2 - p_5 - p_{31})^2} \gamma_{\mu} \frac{\not\!p_2 - \not\!p_5 + m_c}{(p_2 - p_5)^2 - m_c^2} \not\! \epsilon (p_2) v_{s}(p_5),
\nonumber \\
i T_{18} &=& ieg^3Q_b \mathcal{C}_{18} \bar{u}_{s^{\prime}}(p_4) \gamma^{\mu} \frac{\not\!p_1 - \not\!p_{32} + m_b}{(p_1 - p_{32})^2 - m_b^2} \not\! \epsilon (p_1) \frac{\Pi(p_3)}{(p_2 - p_5 - p_{31})^2} \gamma_{\mu} \frac{\not\!p_2 - \not\!p_5 + m_c}{(p_2 - p_5)^2 - m_c^2} \not\! \epsilon (p_2) v_{s}(p_5),
\nonumber \\
i T_{19} &=& ieg^3Q_b \mathcal{C}_{19} \bar{u}_{s^{\prime}}(p_4) \not\! \epsilon (p_1) \frac{\not\!p_4 - \not\!p_1 + m_b^2}{(p_1 - p_4)^2 - m_b^2} \gamma^{\mu} \frac{\Pi(p_3)}{(p_2 - p_5 - p_{31})^2} \not\! \epsilon (p_2) \frac{\not\!p_{31}- \not\!p_2 + m_c}{(p_2 - p_{31})^2 - m_c^2} \gamma_{\mu} v_{s}(p_5),
\nonumber \\
i T_{20} &=& ieg^3Q_b \mathcal{C}_{20} \bar{u}_{s^{\prime}}(p_4) \gamma^{\mu} \frac{\not\!p_1 - \not\!p_{32} + m_b}{(p_1 - p_{32})^2 - m_b^2} \not\! \epsilon (p_1) \frac{\Pi(p_3)}{(p_2 - p_5 - p_{31})^2} \not\! \epsilon (p_2) \frac{\not\!p_{31}- \not\!p_2 + m_c}{(p_2 - p_{31})^2 - m_c^2} \gamma_{\mu} v_{s}(p_5),\nonumber 
\end{eqnarray}
\begin{eqnarray}
i T_{21} &=& ieg^3Q_b \mathcal{C}_{21} \bar{u}_{s^{\prime}}(p4) \not\! \epsilon (p_1) \frac{\not\!p_4 - \not\!p_1 + m_b}{(p_1 - p_4)^2 - m_b^2} \gamma_{\rho} \frac{\Pi(p_3)}{(p_{31} + p_5)^2 (p_{31} + p_5 - p_2)^2} \gamma_{\nu} \nonumber \\
&& \cdot [g^{\mu \nu}(p_2 + p_5 + p_{31})^{\rho} + g^{\nu \rho}(p_2 - 2p_{31} -2p_5 )^{\mu} + g^{\rho \mu}(p_5 + p_{31} - 2p_2)^{\nu} ] \epsilon_{\mu} v_{s}(p_5),
\nonumber \\
i T_{22} &=& ieg^3 Q_b\mathcal{C}_{22} \bar{u}_{s^{\prime}}(p4) \gamma_{\rho} \frac{\not\!p_1 - \not\!p_{32} + m_b }{(p_1 - p_{32})^2 - m_b^2} \not\!\epsilon (p_1) \frac{\Pi(p_3)}{(p_{31} + p_5)^2 (p_{31} + p_5 - p_2)^2} \gamma_{\nu} \nonumber \\
&& \cdot [g^{\mu \nu}(p_2 + p_5 + p_{31})^{\rho} + g^{\nu \rho}(p_2 - 2p_{31} -2p_5 )^{\mu} + g^{\rho \mu}(p_5 + p_{31} - 2p_2)^{\nu} ] \epsilon_{\mu} v_{s}(p_5),
\nonumber \\
i T_{23} &=& -ieg^3Q_c \mathcal{C}_{23} \bar{u}_{s^{\prime}}(p4) \gamma_{\nu} \frac{\Pi(p_3)}{(p_{32} + p_4)^2 (p_{32} + p_4 - p_2)^2} \gamma_{\rho} \frac{\not\!p_1 - \not\!p_5 + m_c}{(p_1 - p_5)^2 - m_c^2} \not\!\epsilon (p_1) \nonumber \\
&& \cdot [g^{\mu \nu}(p_2 + p_4 + p_{32})^{\rho} + g^{\nu \rho}(p_2 - 2p_4 - 2p_{32})^{\mu} + g^{\rho \mu}(p_4 + p_{32} - 2p_2)^{\nu}] \epsilon_{\mu} v_{s}(p_5),
\nonumber \\
i T_{24} &=& -ieg^3 Q_c\mathcal{C}_{24} \bar{u}_{s^{\prime}}(p4) \gamma_{\nu} \frac{\Pi(p_3)}{(p_{32} + p_4)^2 (p_{32} + p_4 - p_2)^2} \not\!\epsilon (p_1) \frac{\not\!p_{31} - \not\!p_1 + m_c}{(p_1 - p_{31})^2 - m_c^2} \gamma_{\rho} \nonumber \\
&& \cdot [g^{\mu \nu}(p_2 + p_4 + p_{32})^{\rho} + g^{\nu \rho}(p_2 - 2p_4 - 2p_{32})^{\mu} + g^{\rho \mu}(p_4 + p_{32} - 2p_2)^{\nu}] \epsilon_{\mu} v_{s}(p_5).\nonumber 
\end{eqnarray}

The four hard scattering amplitudes $T_j$ ($T= {\sum_{j=1}^{4}} {T_j}$) for the subprocess $\gamma(p_1)+c(p_2)\to B_c^{**}(p_3)+b(p_4)$ are
\begin{eqnarray}
i T_{1} &=& ieg^2 Q_c \mathcal{C}'_1 \bar{u}_{s^{\prime}}(p_4) \gamma^{\mu} \frac{\Pi(p_3)}{(p_{32}+p_4)^2} \gamma_{\mu} \frac{\not\!p_3 + \not\!p_4 + m_c}{(p_3 + p_4)^2 - m_c^2} \not\! \epsilon(p_1) u_{s}(p_2), \label{eqmc1}
\nonumber \\
i T_{2} &=& ieg^2 Q_b \mathcal{C}'_2 \bar{u}_{s^{\prime}}(p_4) \not\! \epsilon(p_1) \frac{\not\!p_4 - \not\!p_1 + m_b}{(p_1 - p_4)^2 - m_b^2} \gamma^{\mu} \frac{\Pi(p_3)}{(p_2-p_{31})^2} \gamma_{\mu} u_{s}(p_2), \label{eqmc2}
\nonumber \\
i T_{3} &=& i eg^2 Q_b\mathcal{C}'_3 \bar{u}_{s^{\prime}}(p_4) \gamma^{\mu} \frac{\not\!p_1 - \not\!p_{32} + m_b}{(p_1 - p_{32})^2 - m_b^2} \not\! \epsilon(p_1) \frac{\Pi(p_3)}{(p_2-p_{31})^2} \gamma_{\mu} u_{s}(p_2), \label{eqmc3}
\nonumber \\
i T_{4} &=& i eg^2 Q_c\mathcal{C}'_4 \bar{u}_{s^{\prime}}(p_4) \gamma^{\mu} \frac{\Pi(p_3)}{(p_{32}+p_4)^2}\not\! \epsilon(p_1) \frac{\not\!p_2 - \not\!p_4 - \not\!p_{32} + m_c}{(p_2 - p_4 - p_{32})^2 - m_c^2} \gamma_{\mu} u_{s}(p_2),\label{eqmc4}\nonumber 
\end{eqnarray}
and the amplitudes for the subprocess $\gamma(p_1)+\bar{b}(p_2)\to B_c^{**}(p_3)+\bar{c}(p_4)$ can be easily obtained from that of  $\gamma(p_1)+c(p_2)\to B_c^{**}(p_3)+b(p_4)$. There $\Pi(p_3)$ stands for the spin-projection operator that depicts the $(c\bar{b})$-pair evolving into the $B^{**}_c$ meson
\begin{equation*}
\Pi(p_3) = \sqrt{M} \left ( \frac{\frac{m_b}{M} \not\!p_3 - \not\!q -m_b }{2m_b} \right ) \Gamma \left( \frac{ \frac{m_c}{M} \not\!p_3 + \not\!q + m_c}{2m_c} \right),
\end{equation*}
where $\Gamma = \gamma^5$ for the spin-singlet state, and $\Gamma= \gamma^{\beta}$ for the spin-triplet state. $\epsilon(p_1)$ and $\epsilon(p_2)$ are polarization vectors of the initial photon and gluon. $p_{31}$ and $p_{32}$ are four-momenta of the $c$-quark and $\bar{b}$-quark in the $B^{**}_c$ meson, $p_{31} = \frac{m_c}{M}p_3+q$ and $p_{32} = \frac{m_b}{M}p_3-q$. The overall color factor $\mathcal{C}_l$ and $\mathcal{C}^{\prime}_l$ are given by
\begin{eqnarray}
\mathcal{C}_{l=(3,5,6,9,11,12,15,16,19,20)} &=& \frac{1}{\sqrt{3}}(T^BT^AT^B)_{\alpha \beta},~~~~~\nonumber \\
\mathcal{C}_{l=(1,2,4,7,8,10,13,14,17,18)}&=& \frac{1}{\sqrt{3}}(T^AT^BT^B)_{\alpha \beta},~~~~~\nonumber \\
\mathcal{C}_{l=(21,...,24)} &=& \frac{i}{\sqrt{3}}f^{ABC}(T^BT^C)_{\alpha \beta},~~~~~\nonumber \\
\mathcal{C}'_{l=(1,...,4)} &=& \frac{1}{\sqrt{3}}C_F\delta_{\lambda \alpha},~~~~~\nonumber
\end{eqnarray}
where the factor $\frac{1}{\sqrt{3}}$ is due to the color-singlet nature of $B^{**}_c$ meson. Superscript $A$ stands for the color of the incident gluon, $\alpha$ and $\beta$ are the color indices of the final outgoing $b$-quark and $\bar{c}$-quark,  and $\lambda$ is the color index of initial incoming $c$ quark. $C_F=\frac{4}{{3}}$ is one of the Casimir operator eigenvalues of SU($3$), $f^{ABC}$ are the structure constants of SU($3$) and $T^{A(B,C)}$ is the SU($3$) generator in the fundamental representation. $Q_b=-\frac{1}{3}$  and $Q_c=\frac{2}{3}$  are electric charges of the $b$-quark and $c$-quark, respectively.

\end{document}